\documentstyle[twocolumn,aps,prl,epsf]{revtex}
\def\d{{\rm d}}
\def\i{{\rm i}}
\def\res#1#2#3{$\!\!(#1\pm #2)\cdot 10^{#3}\!\!$}
\def\bom#1{{\mbox{\boldmath $#1$}}}

\def\a#1#2{\la#1#2\ra}
\def\b#1#2{[#1#2]}
\def\s#1#2{s_{#1#2}}
\def\c#1#2#3#4{\la#1|(#2+#3)|#4\ra}
\def\t#1#2#3{t_{#1#2#3}}
\def\vspaceinarray{\\ \nn &~&~}
\def\nextline{\vspaceinarray\\ \nn && \qquad}
\def\nexteq{\vspaceinarray\vspaceinarray\\ &&}

\newcommand{\la}{\langle}
\newcommand{\ra}{\rangle}
\newcommand\lra{\leftrightarrow}
\def\bra#1{\la{#1}|}
\def\ket#1{|{#1}\ra}
\def\nbra#1{{}_n\la{#1}|}
\def\nket#1{|{#1}\ra_n}
\def\mbra#1{{}_4\la{#1}|}
\def\mket#1{|{#1}\ra_4}
\def\st#1{\lower.75em\hbox{$|_{#1}$} }

\newcommand\as{{\alpha_s}}
\newcommand\ep{{\varepsilon}}
\newcommand\NLO{next-to-leading order }

\newcommand\qb{{\bar q}}
\newcommand\Qb{{\bar Q}}
\newcommand{\Nc}{{N_{\rm c}}}
\newcommand{\NA}{{N_A}}
\newcommand{\Nf}{{N_{\rm f}}}
\newcommand{\Ord}{{\rm O}}
\newcommand\M{{\cal M}}

\newcommand\gt{{\tilde g}}

\newcommand\D{{\cal D}}
\newcommand\Tr{{\rm Tr}}
\newcommand{\beq}{\begin{equation}}
\newcommand{\eeq}{\end{equation}}
\newcommand{\beeq}{\begin{eqnarray}}
\newcommand{\eeeq}{\end{eqnarray}}
\newcommand\nn{\nonumber}
\newcommand\ycut{y_{\rm cut}}
\newcommand\yflip{y_4}
\newcommand\yflipD{y_4^D}
\newcommand\yflipC{y_4^C}

\begin{document}

\twocolumn[\hsize\textwidth\columnwidth\hsize\csname      
@twocolumnfalse\endcsname

\preprint{hep-ph/9806317}
\title{Next-to-leading order calculation of four-jet observables in
electron-positron annihilation}
\author{Zolt\'an Nagy}
\address{Department of Theoretical Physics, KLTE,
H-4010 Debrecen P.O.Box 5, Hungary}
\author{Zolt\'an Tr\'ocs\'anyi}
\address{Institute of Nuclear Research of the Hungarian Academy of Sciences,
H-4001 Debrecen P.O.Box 51, Hungary
\\
and Department of Theoretical Physics, KLTE,
H-4010 Debrecen P.O.Box 5, Hungary}
\date{\today}
\maketitle

\begin{abstract}
The production of four jets in electron-positron annihilation allows for
measuring the strong coupling and the underlying group structure of the
strong interaction simultaneously. This requires next-to-leading order
perturbative prediction for four-jet observables. In this paper we
describe the theoretical formalism of such a calculation with
sufficient details. We use the dipole method to construct a Monte Carlo
program that can be used for calculating any four-jet observable at the
\NLO accuracy. As new results, we present the \NLO prediction for the
thrust minor, $\yflip$ and C parameter (at C $\ge$ 0.75) four-jet shape
variables and the four-jet rates with the Cambridge jet clustering
algorithm. 

\end{abstract}
]

\section{Introduction}

Electron-positron annihilation into hadrons is the cleanest process to
test Quantum Chromodynamics (QCD) \cite{QCD} in high energy elementary
particle reactions. In this process the initial state is completely
known and there is a lot of quantities, for instance the total cross
section and jet related correlations, that depend on the long distance
properties of the theory very little. These quantities can be calculated
in perturbative QCD as a function of a single parameter, the strong
coupling. For this reason the various QCD tests at electron-positron
colliders \cite{KEK,SLAC,ALEPH,DELPHI,OPAL,L3} can be regarded as
experiments for determining $\as$.

The other ingredient of QCD, that is in principle free, is the underlying
gauge group.  Although by now nobody questions that QCD is based upon SU(3)
gauge theory, the ``full'' measurement of QCD, that is the simultaneous
measurement of the strong coupling and the eigenvalues of the quadratic
Casimirs of the underlying gauge theory, the $C_F$ and $C_A$ color
charges, is not a purely academic exercise. The possible existence of
light gluinos \cite{lg} influences both the value of $\as$ and the
measured value of the color charges (or, assuming SU(3)$_{\rm c}$, the
value of the light fermionic degrees of freedom $\Nf$). Thus the only
consistent framework to check whether the data favor or exclude the
additional degrees of freedom is a simultaneous fit of these parameters
to data.

In principle any observable depends on these basic parameters. The
sensitivity of a given observable on the color charges however, is
influenced by the fact that in perturbation theory the three gluon
coupling appears at tree level first for four-jet final states.
In the total cross section and for three-jet like quantities the adjoint
color charge appears only in the radiative corrections. Therefore,
four-jet observables seem to be the best candidates to measure the color
factors. Indeed, during the first phase of operation of the Large
Electron Positron Collider (LEP) four-jet events were primarily used
for measuring $C_F$ and $C_A$ \cite{Cmeasure}.%\cite{L3,a34,DELPHI,OPAL,ALEPH}.
These measurements however, were not complete in the sense mentioned
above. The lack of knowledge about the perturbative prediction for
four-jet observables at O($\alpha_s^3$) prevented the experimental
collaborations from fixing the absolute normalization of the perturbative
prediction, therefore, $\as$ could not be measured using the same
observables.

Recently the \NLO corrections to various four-jet observables have been
calculated \cite{SDjets,DSjets,Signer,NTDpar,NTFox,NTangulars,Glover}.
In this article we give sufficient details of our work
\cite{NTDpar,NTFox,NTangulars} and present several new results for
\NLO predictions of four-jet observables that were not published before.
The important development that made possible these calculations was that
the one-loop amplitudes for the relevant QCD subprocesses, i.e.\ for 
$e^+e^-\to$ 4 partons, became available. In Refs.~\cite{GM4q,CGM2q2g}
Campbell, Glover and Miller introduced FORTRAN programs that calculate
the \NLO squared matrix elements of the $e^+e^- \to \gamma^* \to \qb q \Qb Q$
and $\qb q g g$ processes.  In Refs.~\cite{BDKW4q,BDK2q2g} Bern, Dixon,
Kosower and Wienzierl gave analytic formulas for the helicity
amplitudes of the same processes with the $e^+e^- \to Z^0 \to$ 4
partons channel included as well. For the sake of completeness, in our
work we use the amplitudes of Refs.~\cite{BDK2q2g} for the loop
corrections. Although the tree-level helicity amplitudes for the
$e^+e^-\to$ 5 partons subprocesses had been known \cite{a5parton}, we
calculated them anew and present the results in this article in terms
of Weyl spinors conforming with the notation used for describing the
one-loop helicity amplitudes \cite{BDK2q2g}. We also present the
previously unpublished color linked helicity dependent Born matrix
elements for the $e^+e^-\to$ 4 partons processes, which are needed also
for the \NLO calculation of the three-jet production in deep inelastic
scattering and for vector boson plus two-jet production in hadron
collisions.

%The rest of the paper is organized as follows.
In Sec.~II we give
details of the analytical and numerical calculation and describe how we
parametrise our results. In Sec.~III we present the complete
O($\alpha_s^3$) predictions for the four-jet rates using the Durham
algorithm \cite{durham} and the Cambridge algorithm proposed recently
\cite{cambridge} and make a comparison of these algorithms from the
perturbation theory point of view. We show the \NLO results for the
shape variables thrust minor and $y_4$ often used in experimental
analyses and for the C parameter distribution at C $\ge$ 0.75. Sec.~IV
contains our conclusions. We present analytic results for the four- and
five-parton tree-level helicity amplitudes in Appendix A,
and perform the color summation in Appendix B.
%for the
%color-correlated four-parton Born-level matrix elements and for the
%four-, five-parton Born-level matrix elements in Appendix B. 

\section{Details of the calculation}

\subsection{Cancellation of infrared divergences}

It is well-known that the \NLO correction is a sum of two integrals,
\beq
\label{sNLO}
\sigma^{\rm NLO} \equiv \int \d\sigma^{\rm NLO}
= \int_5 \d\sigma^{\rm R}
+ \int_4 \d\sigma^{\rm V}\:,
\eeq
where $\d\sigma^{\rm R}$ is an exclusive cross section of five partons
in the final state:
\beq
\int_5 \d\sigma^{\rm R}
= \int \d\Gamma^{(5)} <|\M_5^{\rm tree}|^2> J_5\:,
\eeq
and $\d\sigma^{\rm V}$ is the one-loop correction to the process with
four partons in the final state:
\beq
\int_4 \d\sigma^{\rm V}
= \int \d\Gamma^{(4)} <|\M_4^{\rm 1-loop}|^2> J_4 \:,
\eeq
--- the real and virtual corrections. Although we specify our
formulas to the case of four-jet calculation, one can use the
formulas of this section to obtain the $m$-jet cross section by simply
changing 4 (5) to $m\;(m+1)$.

The two integrals on the
right-hand side of Eq.~(\ref{sNLO}) are separately divergent in $d=4$
dimensions, but their sum is finite provided the jet function $J_n$
defines an infrared safe quantity, which formally means that
\beq
J_5 \to J_4
\eeq
in any case, where the five-parton and the four-parton configurations
are kinematically degenerate. The presence of singularities means that
the separate pieces have to be regularized, and the divergences have to
be cancelled. We use dimensional regularization in $d=4-2\ep$ dimensions
\cite{dimreg}, in which case the divergences are replaced by double and
single poles in $\ep$. We assume that ultraviolet renormalization of
all Green functions to one-loop order has been carried out, so the
poles are of infrared origin. In order to obtain the finite sum, we use
a slightly modified version of the dipole method of Catani and Seymour
\cite{CSdipole} that exposes the cancellation of the infrared
singularities directly at the integrand level.

The reason for modifying the original dipole formalism is numerical.
The essence of the dipole method is to define a single subtraction term
$\d\sigma^{\rm A}$ --- the dipole subtraction ---, that reguralizes the
real correction in all of its singular (soft and collinear) limits. Thus,
the two singular integrals in Eq.~(\ref{sNLO}) are substituted by two
finite ones:
\beq
\label{sNLO2}
\sigma^{\rm NLO}
= \int_5 \d\sigma_5^{\rm NLO} + \int_4 \d\sigma_4^{\rm NLO}\:,
\eeq
where
\beq
\label{r-a}
\int_5 \d\sigma_5^{\rm NLO}
= \int_5 \left[\d\sigma^{\rm R}-\d\sigma^{\rm A}\right]
\eeq
and
\beq
\int_4 \d\sigma_4^{\rm NLO}
= \int_4 \left[\d\sigma^{\rm V}+\int_1 \d\sigma^{\rm A}\right]
\eeq
There are many ways to define the subtraction term, but all
must lead to the same finite \NLO correction. Since the virtual
correction is not positive definite, then depending on the size of the
subtraction term it may happen that either $\d\sigma_5^{\rm NLO}$ or
$\d\sigma_4^{\rm NLO}$ is not positive definite. From numerical point of
view the best situation is when both are positive definite, so that
numerical cancellation of terms with opposite sign does not occur.
We define the subtraction term as a function of a parameter $\alpha \in
(0,1]$ which essentially controls the region of the five-parton phase
space over which the subtraction is non-zero such that $\alpha=1$ means
the full dipole subtraction (see subsection II.B). By tuning the value of
$\alpha$, we can achieve that we add two positive definite integrals for
almost all values of the observable to
obtain the full correction. We use an $\alpha \simeq 0.1$, which is
advantageous also for saving CPU time: The large number of dipole
terms and their somewhat complicated analytic structure makes the
evaluation of the subtraction term rather time consuming. Constraining
the phase space over which the subtraction is zero we can speed up the
program.

In spite that the five-parton integral in Eq.~(\ref{r-a}) is finite,
we introduce a very small cutoff in the phase space around
the singular regions. Such a cutoff does not alter the value of the
integral, but helps avoiding the cancellation of very large numbers that
could lead to arbitrary values close to the singularity due to the finite
machine precision.  This cutoff is useful, but is also dangerous: if
the subtraction is not correct, the five-parton integral becomes
finite, but incorrect. The third advantage of using the parameter
$\alpha$ is that such errors can be spotted by varying the value of
$\alpha$ and checking whether the full correction is independent of 
this parameter.

\subsection{Dipole formulas for final state singularities}

In this subsection we recall those dipole factorization formulas that
are relevant to our calculation. We do this only to the extent that we
can define the simple modification to the original formalism
and the explicit cross section formulas of our calculation unambiguously.
For further details we refer to the original work of Catani and Seymour
\cite{CSdipole}.

In the dipole method the subtraction term is a sum of several dipole
terms,
\beq
\label{dsA}
\int_5 \d\sigma^{\rm A}
= \sum_{k\ne i,j}\int\! \d\Gamma^{(5)}\, \D_{ij,k}\, J_4\:,
\eeq
where the dipole $\D_{ij,k}$ is a function of the final state momenta
$p_l$ and is given by
\beeq
\label{dipole}
&&\D_{ij,k}(p_1,\ldots,p_5) =
- \frac{1}{2 p_i \cdot p_j}
\\ \nn &&\;
\times\mbra{\ldots, {\widetilde {ij}},\ldots, {\widetilde k},\ldots}
\,\frac{{\bom T}_k \cdot {\bom T}_{ij}}{{\bom T}_{ij}^2}
\; {\bom V}_{ij,k} \, 
\mket{\ldots, {\widetilde {ij}},\ldots, {\widetilde k},\ldots} \:.
\eeeq 
In Eq.~(\ref{dipole})
$\mket{\ldots, {\widetilde {ij}},\ldots, {\widetilde k},\ldots}$
is a vector in color + helicity space defining the four-parton amplitude
obtained from the original five-parton Born amplitude by replacing $a)$ the
partons $i$ and $j$ with a single parton ${\widetilde {ij}}$ ({\em the
emitter\/}) and $b)$ the parton $k$ with the parton $\widetilde k$
({\em the spectator\/}). The momenta of the spectator and the emitter
are given in terms of a dimensionless variable 
\beq
y_{ij,k} = \frac{p_ip_j}{p_ip_j+p_jp_k+p_kp_i}\:,
\eeq
as
\beq
\label{dipolemom}
{\widetilde p}_k^\mu = \frac{1}{1-y_{ij,k}} \,p_k^\mu \:,\quad
{\widetilde p}_{ij}^\mu =
p_i^\mu  + p_j^\mu  - \frac{y_{ij,k}}{1-y_{ij,k}} \,p_k^\mu\:.
\eeq
${\bom T}_k$ and ${\bom T}_{ij}$ are the color charges of
the spectator and the emitter. These color charges are defined by their
action onto the color space: If particle $i$ emits a gluon with color
index $c$ then the color-charge operator ${\bom T}_i$ has the following
matrix element in color space 
\beeq
&&
\bra{c_1,\ldots, c_i,\ldots, c_4, c} {\bom T}_i
\ket{b_1,\ldots, b_i,\ldots, b_4} =
\\ \nn && \qquad\qquad
\delta_{c_1 b_1}\ldots T_{c_i b_i}^c\ldots\delta_{c_4 b_4} \:, 
\eeeq
where $T_{c_ib_i}^c \equiv F_{c_ib_i}^c = -\i f_{cc_ib_i}$
(color-charge matrix in the adjoint representation)  if the emitting
particle $i$ is a gluon and $T_{c_ib_i}^c \equiv t^c_{c_ib_i}$
(color-charge matrix in the fundamental representation) if the emitting
particle $i$ is a quark (in the case of an antiquark emitter
$T_{c_ib_i}^c \equiv {\bar t}^c_{c_ib_i} = - t^c_{b_ic_i }$).

In Eq.~(\ref{dipole}) the splitting matrices, ${\bom V}_{ij,k}$ are
matrices in the helicity space of the emitter. They depend on the
kinematic variables $y_{ij,k}$ and 
\beq
{\tilde z}_i =
\frac{p_i{\widetilde p}_k}{{\widetilde p}_{ij}{\widetilde p}_k}
\eeq
and take different forms for the splitting of different partons
(see Eqs.~(5.7--5.9) in Ref.~\cite{CSdipole}).

The definition of the dipole momenta makes possible the exact
factorization of the five-particle phase space into a four-particle and
a one-particle phase space
\beeq
&&
\d\Gamma^{(5)} (p_1,\ldots,p_{5})=
\\ \nn && \qquad
\d\Gamma^{(4)} (\ldots,{\widetilde p}_{ij},\ldots,{\widetilde p}_k,\ldots)
\,\left[\d p_i({\widetilde p}_{ij},{\widetilde p}_k)\right]\:,
\eeeq
where
\beq
\label{dpi}
\left[ \d p_i({\widetilde p}_{ij},{\widetilde p}_k) \right] 
= \frac{\d^{d}p_i}{(2\pi)^{d-1}} \,\delta_+(p_i^2) \,
{\cal J}(p_i;{\widetilde p}_{ij},{\widetilde p}_k) \:,
\eeq 
and the Jacobian factor is
\beq
\label{Jac}
{\cal J}(p_i;{\widetilde p}_{ij},{\widetilde p}_k)
= \Theta(1- {\tilde z}_i) \,\Theta(1-y_{ij,k}) \,
\frac{(1-y_{ij,k})^{d-3}}{1- {\tilde z}_i} \:.
\eeq
As mentioned before, we modify the original formalism such that we
constrain the phase space over which the dipoles are subtracted:
\beq
\label{dsAmod}
\int_5 \d\sigma^{\rm A}(\alpha)
= \sum_{k\ne i,j}\int\!\d\Gamma^{(5)}\:\D_{ij,k}\:\Theta(y_{ij,k}<\alpha)
\:J_4\:,
\eeq
with $\alpha\in(0,1]$.
The jet function in Eq.~(\ref{dsAmod}) is a four-parton jet function that
depends on the momenta ${\widetilde p}_{ij}$ and ${\widetilde p}_k$, but
not on $p_i$, therefore, the integral over the one-parton phase space can
be performed analytically. It can be shown that after integration of
the dipole ${\cal D}_{ij,k}(p_1,\ldots,p_{5})$ over
$[\d p_i({\widetilde p}_{ij},{\widetilde p}_k)]$, only color
correlations survive \cite{CSdipole}, in the form:
\beeq
\label{intdip}
&&\int\!\left[\d p_i({\widetilde p}_{ij},{\widetilde p}_k)\right]\,
\D_{ij,k}(p_1,\ldots,p_{5})\, \Theta(y_{ij,k}<\alpha) =
\\ \nn && \qquad\qquad
- \,{\cal V}_{ij,k}(\alpha) \:
\frac{1}{{\bom T}_{ij}^2}
\left|\M^{ij,k}_4
(\ldots,{\widetilde {ij}},\ldots, {\widetilde k},\ldots)\right|^2\:,
\eeeq
where
\beq
\label{Mij}
\left|\M^{i,j}_4 (1,\ldots,4)\right|^2 =
\mbra{1,\ldots,4} \,{\bom T}_i \cdot {\bom T}_j \, \mket{1,\ldots,4} 
\eeq
are the color-correlated four-parton tree matrix elements. Their explicit
expressions are given in Appendix B. In Eq.~(\ref{intdip})
\beeq
\label{vijep}
&&
{\cal V}_{ij,k}(\alpha) =
\\ \nn &&\quad
\int\!\left[\d p_i({\widetilde p}_{ij},{\widetilde p}_k)\right]
\,\Theta(y_{ij,k}<\alpha)
\, \frac{1}{2 p_i \cdot p_j} \; <{\bom V}_{ij,k}>
\,\equiv 
\\ \nn &&\quad
\frac{\as}{2\pi}
\frac{1}{\Gamma(1-\ep)} \left(
\frac{4\pi \mu^2}{2 {\widetilde p}_{ij}{\widetilde p}_k} \right)^{\ep}
{\cal V}_{ij}(\ep,\alpha) \;,
\eeeq
where $<{\bom V}_{ij,k}>$ denotes the average of
${\bom V}_{ij,k}$ over the polarisations of the emitter parton
${\widetilde {ij}}$.
The functions ${\cal V}_{ij}(\ep,\alpha)$ depend only on the flavour
indices $i$ and $j$. Rewriting the one-particle phase space in terms of
the kinematic variables ${\tilde z}_i$ and $y$, from the definition
of ${\cal V}_{ij}(\ep,\alpha)$ in Eq.~(\ref{vijep}) we obtain
\twocolumn[\hsize\textwidth\columnwidth\hsize\csname      
@twocolumnfalse\endcsname
\beq
\label{intvij}
{\cal V}_{ij}(\ep,\alpha) =
\int_0^1\!\d{\tilde z}_i \,
\left({\tilde z}_i (1-{\tilde z}_i) \right)^{-\ep}
\int_0^\alpha\!\d y \, y^{-1-\ep} \left( 1-y \right)^{1-2\ep} \;
\frac{<{\bom V}_{ij,k}({\tilde z}_i;y)>}{8 {\pi} \as
\mu^{2\ep}} \;,
\eeq
where the spin-averaged splitting functions are given in Eqs.~(5.29-5.31)
of Ref.~\cite{CSdipole}.  Performing the integration in
Eq.~(\ref{intvij}), we find
\beeq
\label{vqgep}
&&
{\cal V}_{qg}(\ep,\alpha) 
= C_F \left\{ \left(\frac{1}{\ep^2} -\log^2 \alpha \right)
+ \frac{3}{2}\left(\frac{1}{\ep}-1+\alpha-\log\alpha \right)
+ 5 - \frac{\pi^2}{2}
+ \Ord(\ep)  \right\} \:,
\\ &&
\label{vqqep}
{\cal V}_{q{\bar q}}(\ep,\alpha) 
= T_R \left\{-\frac{2}{3}\left(\frac{1}{\ep}-1+\alpha-\log\alpha \right)
- \frac{16}{9}
+ \Ord(\ep) \right\} \:,
\\ &&
\label{vggep}
{\cal V}_{gg}(\ep,\alpha) 
= 2 C_A \left\{ \left(\frac{1}{\ep^2} -\log^2\alpha \right)
+ \frac{11}{6}\left(\frac{1}{\ep}-1+\alpha-\log\alpha \right)
+ \frac{50}{9} - \frac{\pi^2}{2}
+ \Ord(\ep)  \right\} \:.
\eeeq
]
It was shown in Ref.~\cite{CSdipole} that after integrating over the
factorized one-particle phase space, the subtraction term can be recast
in the form 
\beq
\label{dsAf}
\int_5 \d\sigma^{\rm A} =
\int_4 \d\Gamma^{(4)}\mbra{1,\ldots,4}\:{\bom I}(\ep,\alpha)\:\mket{1,\ldots,4}
\:J_4\:,
\eeq                           
where the insertion operator ${\bom I}(\ep,\alpha)$ depends on the color
charges and momenta of the four final-state partons in
$\mket{1,\ldots,4}$:
\beeq            
\label{iee}
&&
{\bom I}(p_1,p_2,p_3,p_4;\ep,\alpha) =
- \frac{\as}{2\pi} \frac{1}{\Gamma(1-\ep)}
\\ \nn && \qquad
\times
\sum_{i=1}^4 \;\frac{1}{{\bom T}_{i}^2} \:
{\cal V}_i(\ep,\alpha) \: \sum_{k \neq i} {\bom T}_i \cdot {\bom T}_k
\: \left( \frac{4\pi \mu^2}{2 p_i\cdot p_k} \right)^{\ep} \:.
\eeeq
The singular factors ${\cal V}_i(\ep,\alpha)$, are defined as
\beeq
&&
{\cal V}_{q({\bar q})}(\ep,\alpha) \equiv {\cal V}_{qg}(\ep,\alpha)
%\:,\quad
\\ &&
{\cal V}_g(\ep,\alpha) \equiv \frac{1}{2} {\cal V}_{gg}(\ep,\alpha) + \Nf \,
{\cal V}_{q{\bar q}}(\ep,\alpha)\:.
\eeeq
Using Eqs.~(\ref{vqgep}--\ref{vggep}) they can be written in the
following explicit form:
\beeq
\label{calvexp}
&&
{\cal V}_{i}(\ep,\alpha) =
\\ \nn && \qquad
{\bom T}_{i}^2 \left(\frac{1}{\ep^2} - \frac{\pi^2}{3} - \log^2\alpha\right)
+ \gamma_i \:\left(\frac{1}{\ep} + \alpha - \log \alpha \right) + K_i 
\\ \nn && \qquad
+ \Ord(\ep) \:,                               
\eeeq
where the $\gamma_i$ and $K_i$ constants are defined by
\beeq
\gamma_{q({\bar q})}\!=\!\frac{3}{2} C_F \:,\qquad \qquad && \;
K_{q({\bar q})}\!=\!\left( \frac{7}{2} - \frac{\pi^2}{6} \right) C_F \:,
\\ \nn
\gamma_g\!=\!\frac{11}{6} C_A - \frac{2}{3} T_R \Nf \:,&& \;
K_g\!=\!\left( \frac{67}{18} - \frac{\pi^2}{6} \right) C_A
- \frac{10}{9} T_R \Nf \:.
\eeeq

The formal result of the cancellation mechanism discussed in this
subsection is that the \NLO correction is a sum of two finite integrals
as given in Eq.~(\ref{sNLO2}). We would like to mention that although
both integrals are finite, the integrand of the five-parton integral is
in fact divergent, it contains integrable square-root singularities in
the kinematically degenerate region of the five-parton phase space. The
efficient way to integrate such a function is to apply important
sampling. We apply multichannel Monte Carlo integration for this
purpose, but do not consider the details of this technical point in
this article.

\subsection{The general structure of the results}

Once the phase space integrations in Eq.~(\ref{sNLO2}) are carried out,
the \NLO differential cross section for the four-jet observable $O_4$
at a fixed scale $Q$ takes the general form
\beeq
\label{nloxsec}
&&\frac{1}{\sigma_0}\frac{\d \sigma}{\d O_4}(O_4)
= \eta^2 B_{O_4}(O_4) + \eta^3 C_{O_4}(O_4)\:,
\eeeq
where $\eta \equiv \as(Q)\,C_F/\,2\pi$. The renormalization scale dependence of
the cross section is obtained by the substitution
$\eta \to \eta(\mu)\,(1+\beta_0/C_F\,\ln x_\mu)$,
with $\eta(\mu) \equiv \as(\mu)\,C_F/\,2\pi$,
which yields
\beeq
\label{nloxsec2}
&&\frac{1}{\sigma_0}\frac{\d \sigma}{\d O_4}(O_4)
= \eta(\mu)^2 B_{O_4}(O_4)
\\ \nn &&\qquad
+ \eta(\mu)^3
\left[B_{O_4}(O_4) \frac{\beta_0}{C_F}
\ln x_\mu^2 + C_{O_4}(O_4)\right]\:.
\eeeq
In Eq.~(\ref{nloxsec2}) $\sigma_0$ denotes the Born cross section for the
process $e^+e^-\to \qb q$, $\mu$ is the renormalization scale,
$x_\mu = \mu/\!\sqrt{s}$ is the renormalization scale divided by the
total c.m.\ energy and $B_{O_4}$ and $C_{O_4}$ are scale independent
functions, $B_{O_4}$ is the Born approximation and $C_{O_4}$ is the
radiative correction. We use the two-loop expression for the running
coupling,
\beq
\label{twoloopas}
\as(\mu) = \frac{\as(M_Z)}{w(\mu,M_Z)}
\left(
1-\frac{\beta_1}{\beta_0}\frac{\as(M_Z)}{2\pi}
\frac{\ln(w(\mu,M_Z))}{w(\mu,M_Z)}
\right)\:,
\eeq
with
\beq
\label{wqq0}
w(q,q_0) = 1 - \beta_0
\frac{\as(q_0)}{2\pi}\ln\left(\frac{q_0}{q}\right)\:,
\eeq
\beq
\beta_0 = \frac{11}{3}C_A - \frac{4}{3} T_R \Nf\:,
\eeq
\beq
\beta_1 = \frac{17}{3}C_A^2 - 2 C_F T_R \Nf - \frac{10}{3} C_A T_R \Nf \:,
\eeq
with the normalization $T_R=1/2$ in $\Tr(T^aT^{\dag b})=T_R\delta^{ab}$.
The numerical values presented in this letter were obtained at the $Z^0$
peak with $M_Z=91.187$\,GeV, $\Gamma_Z=2.49$\,GeV, $\sin^2\theta_W=0.23$,
$\as(M_Z)=0.118$ and $\Nf=5$ light quark flavors.

In order to make possible the measurement of the color factors, we
write both the Born approximation and the higher order correction as
linear and quadratic forms of ratios of the color charges
\cite{NTloopamps}:
\beq                                                                  
B_4 = B_0 + B_x\,x + B_y\,y \:,
\eeq
and            
\beeq
\label{C4}
&&C_4 = C_0 +\,C_x\,x + C_y\,y + C_z\,z              
\\ \nn && \quad\;\;
+\,C_{xx}\,x^2 + C_{xy}\,x\,y + C_{yy}\,y^2 \:,
\eeeq
where
\beq 
\label{xydef}
x=\frac{C_A}{C_F}\:,\qquad y=\frac{T_R}{C_F}\:.
\eeq
At \NLO the ratio $z$ appears that is related to the square of a cubic
Casimir,
\beq
C_3 = \sum_{a,b,c=1}^\NA \Tr(T^a T^b T^{\dag c}) \Tr(T^{\dag c} T^b
T^a)\:,    
\eeq
via $z=\frac{C_3}{\Nc C_F^3}$. The Born and correction functions $B_i$
and $C_i$ are independent of the underlying gauge group.  In the next
section we present the $B_4$ and $C_4$ functions for various four-jet
observables.

\section{Results}

Four-jet observables can be classified into three major groups:
(i) four-jet rates;
(ii) four-jet event shapes;
(iii) four-jet angular correlations.
%\begin{itemize}
%\item four-jet rates;
%\item four-jet event shapes;
%\item four-jet angular correlations.
%\end{itemize}
Detailed results for observables falling into all three classes were
already presented in the literature. Dixon and Signer gave full account
of the \NLO four-jet rates with three different (E0, Durham and Geneva)
jet algorithms \cite{DSjets}. In Ref.~\cite{NTDpar} we confirmed their
results. Among the four-jet event shapes the D parameter, acoplanarity,
and the Fox-Wolfram moments $\Pi_1$ and $\Pi_4$ were calculated at
O($\alpha_s^3$) in refs.~ \cite{NTDpar} and \cite{NTFox}. The results for
the D parameter were confirmed \cite{DSpriv}. As for angular
correlations Signer presented the leading color corrections in
Ref.~\cite{Signer}, and we added the full corrections in
Ref.~\cite{NTangulars}. In this section we would like to add the
four-jet rate obtained using the Cambridge clustering and several event
shape variables to the list of four-jet observables that are calculated
at the \NLO accuracy. We do not consider the four-jet angular
correlations here.

\subsection{Four-jet rates}

The most important multi-jet observables that are used for determining
the underlying parton structure of hadronic events are the multi-jet
rates. In $e^+e^-$ annihilation the widely known
%JADE \cite{JADE} and
Durham \cite{durham} algorithm have become
indispensable for this purpose. Recently a new jet clustering, the
Cambridge algorithm was proposed as an improved version of the Durham
scheme \cite{cambridge}. This scheme is designed to minimize the formation of
``junk jets'' --- jets formed from hadrons of low transverse
momenta, unconnected to the underlying parton structure. As a result, the
hadronization corrections to the mean jet multiplicities were found
smaller when the Cambridge algorithm is employed than for the Durham
clustering \cite{cambridge}. However, it was shown in
Ref.~\cite{cambridgeB} that the small hadronization corrections found
for the Cambridge algorithm in the study of the mean jet rate are due
to cancellations among corrections for the individual jet production
rates. Apart from the very small values of the resolution parameter,
$\ycut < 10^{-3.2}$, for the individual rates the Durham clustering
shows comparably small (for $\ycut > 10^{-2}$), or even much
smaller hadronization corrections. In this subsection we present the \NLO
production rates for four jets using both algorithms and compare the
size of the radiative corrections.

The four-jet rates are defined as the ratio of the four-jet cross section
to the total hadronic cross section:
\beeq
\label{R4}
&&
R_4=\frac{\sigma_{\rm 4-jet}}{\sigma_{\rm tot}}(\ycut)
\\ \nn &&\quad\;
= \eta^2 B_4(\ycut)
+\eta^3\left(C_4(\ycut)-\frac{3}{2}B_4(\ycut)\right)\:,
\eeeq
where we used $\sigma_{\rm tot}=\sigma_0(1+\frac{3}{2}\eta)$. Setting
the color charges to the SU(3) values, we plot the scale independent
$B_4(\ycut)$ and $C_4(\ycut)$ functions in Figs.~1 and 2 and
tabulate the values for $C_4(\ycut)$ in Table~I.

\begin{figure}
\label{Bycut}
\epsfxsize=8cm \epsfbox{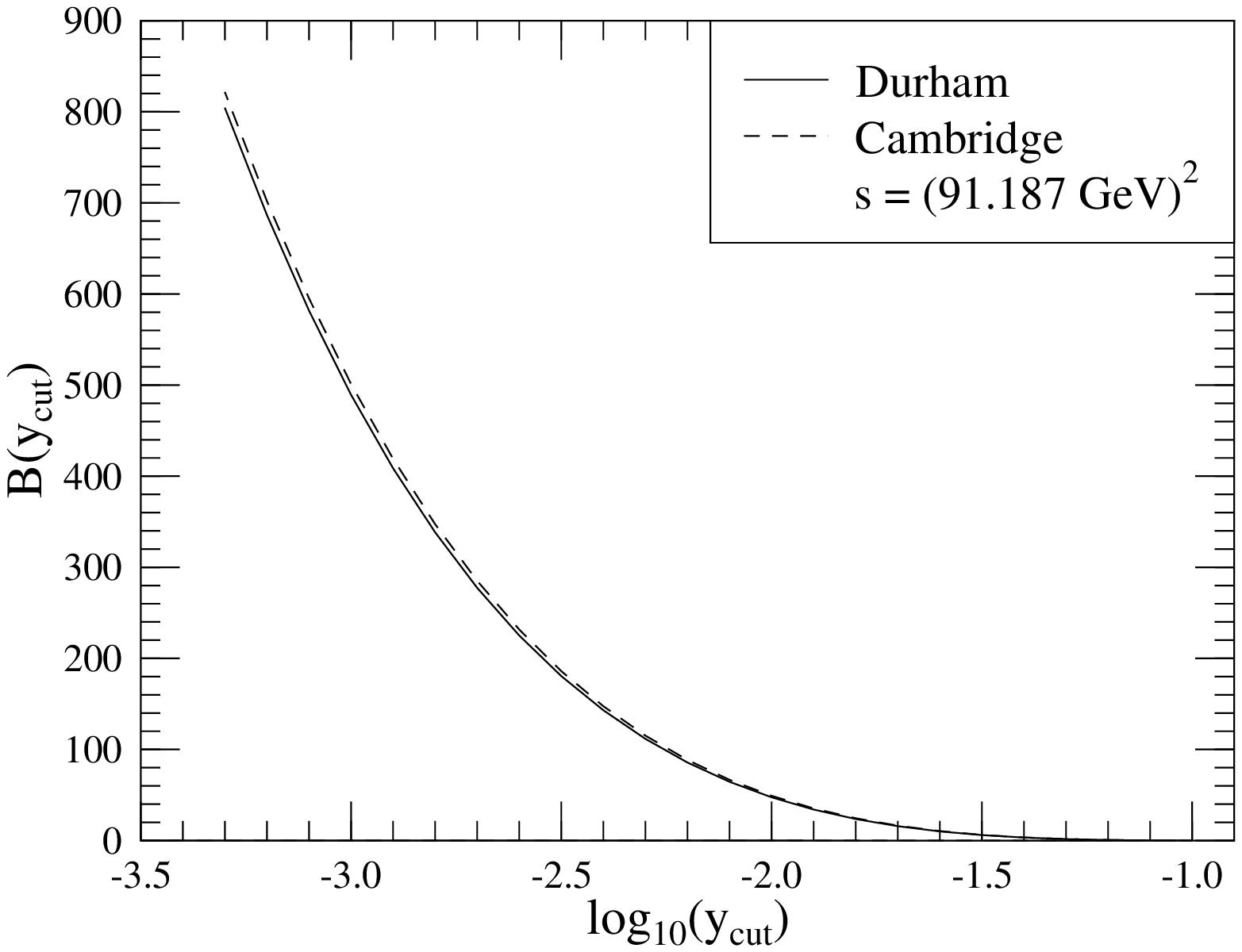}
\caption{The Born function $B_4$ for the four-jet rate as a
function of the resolution variable $\ycut$ with Durham (solid) and
Cambridge (dashed) algorithms. }
\end{figure}
\begin{figure}
\label{Cycut}
\epsfxsize=8cm \epsfbox{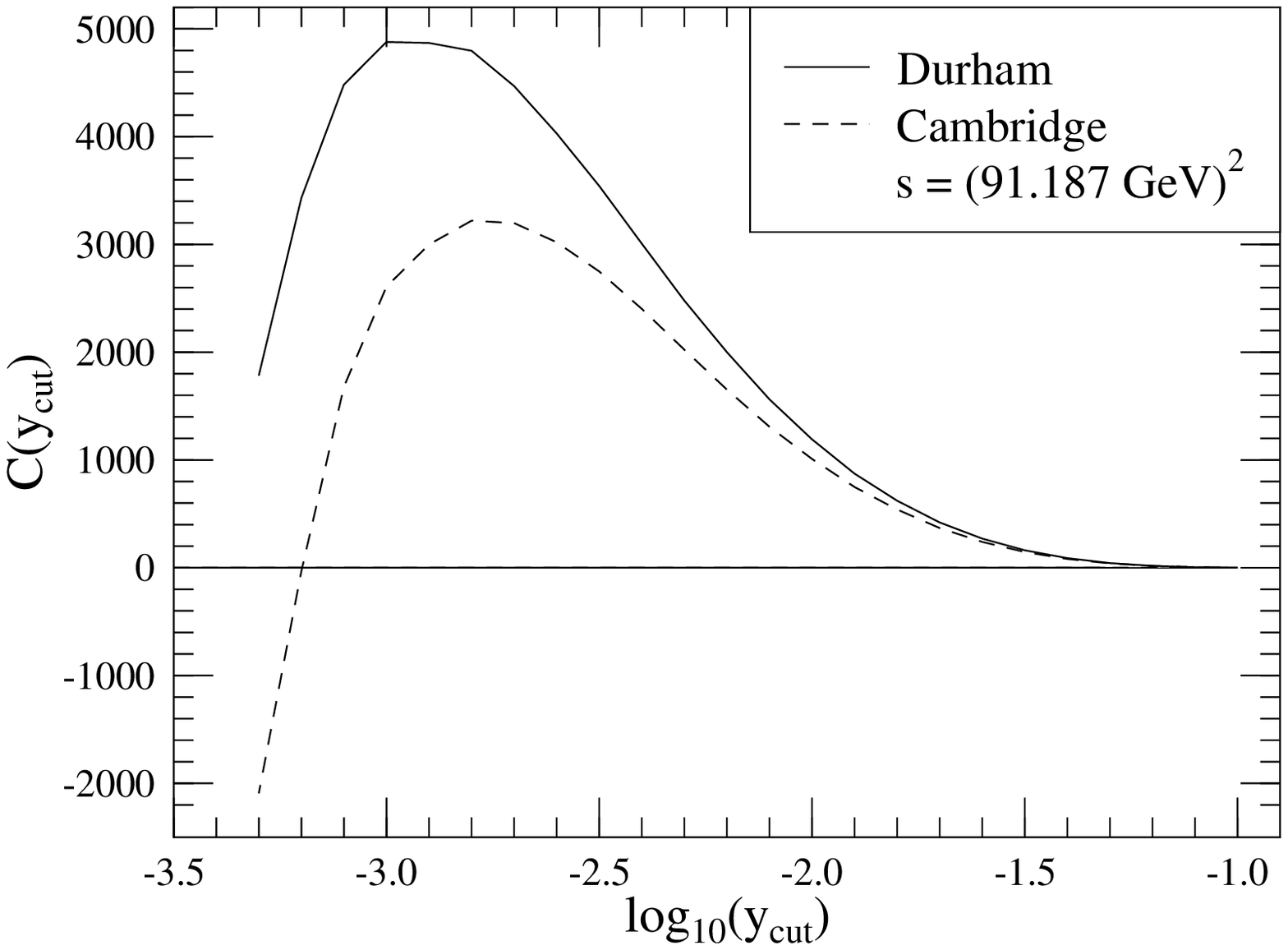}
\caption{The correction function $C_4$ for the four-jet rate as a
function of the resolution variable $\ycut$ with Durham (solid) and
Cambridge (dashed) algorithms. }
\end{figure}

\vbox{\begin{table}             
\caption{Correction functions to the four-jet rates for Durham and
Cambridge algorithms.}
\begin{tabular}{ccc}
  $\log_{10}(y_{cut})$ & $C_{y_{cut}}^{\rm D}$ & $C_{y_{cut}}^{\rm C}$ \\
  \tableline                        
$-0.9$& \res{4.209}{0.655}{-2} & \res{4.375}{0.655}{-2} \\
$-1.0$& \res{9.449}{0.220}{-1} & \res{9.499}{0.230}{-1} \\
$-1.1$& \res{5.411}{0.055}{0} & \res{5.300}{0.057}{0} \\   
$-1.2$& \res{1.769}{0.011}{1} & \res{1.700}{0.012}{1} \\   
$-1.3$& \res{4.321}{0.032}{1} & \res{4.044}{0.033}{1} \\   
$-1.4$& \res{8.893}{0.034}{1} & \res{8.142}{0.038}{1} \\   
$-1.5$& \res{1.619}{0.005}{2} & \res{1.459}{0.006}{2} \\   
$-1.6$& \res{2.705}{0.009}{2} & \res{2.400}{0.010}{2} \\   
$-1.7$& \res{4.201}{0.012}{2} & \res{3.683}{0.014}{2} \\   
$-1.8$& \res{6.221}{0.020}{2} & \res{5.403}{0.021}{2} \\   
$-1.9$& \res{8.730}{0.029}{2} & \res{7.490}{0.032}{2} \\   
$-2.0$& \res{1.191}{0.004}{3} & \res{1.009}{0.005}{3} \\   
$-2.1$& \res{1.563}{0.006}{3} & \res{1.308}{0.007}{3} \\   
$-2.2$& \res{2.000}{0.010}{3} & \res{1.653}{0.010}{3} \\   
$-2.3$& \res{2.478}{0.011}{3} & \res{2.023}{0.012}{3} \\   
$-2.4$& \res{3.007}{0.024}{3} & \res{2.402}{0.025}{3} \\   
$-2.5$& \res{3.542}{0.023}{3} & \res{2.749}{0.027}{3} \\   
$-2.6$& \res{4.029}{0.033}{3} & \res{3.020}{0.036}{3} \\   
$-2.7$& \res{4.469}{0.052}{3} & \res{3.198}{0.063}{3} \\   
$-2.8$& \res{4.797}{0.067}{3} & \res{3.220}{0.077}{3} \\   
$-2.9$& \res{4.869}{0.099}{3} & \res{2.999}{0.108}{3} \\   
$-3.0$& \res{4.878}{0.120}{3} & \res{2.608}{0.132}{3} \\   
$-3.1$& \res{4.482}{0.166}{3} & \res{1.678}{0.178}{3} \\   
$-3.2$& \res{3.430}{0.256}{3} & \res{-3.254}{27.6}{1} \\   
$-3.3$& \res{1.783}{0.300}{3} & \res{-2.093}{0.32}{3} \\  
\end{tabular} 
\end{table}
\noindent

 }

Comparing the values for the two Born functions, we see that at
leading order the Cambridge algorithm gives slightly higher rates and the
difference increases with decreasing $\ycut$. On the other hand, the
correction functions become smaller for Cambridge clustering with
decreasing $\ycut$. The result of these opposite trends is that the
$K$ factors, defined as
\begin{figure}
\label{Kycut}
\epsfxsize=8cm \epsfbox{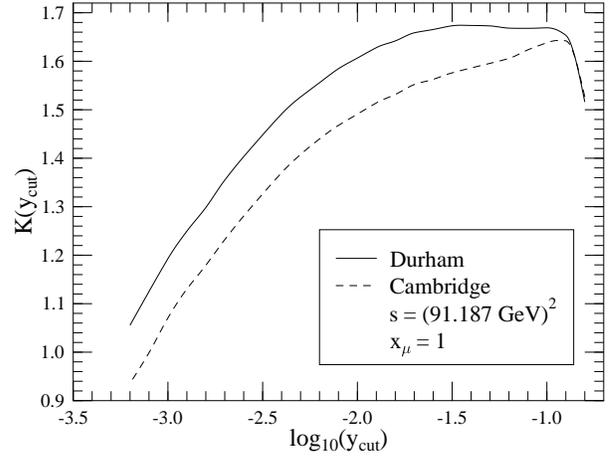}
\caption{$K$ factors as a function of the resolution variable $\ycut$
for Durham (solid) and Cambridge (dashed) algorithms. }
\end{figure}

\beq
K(\ycut) = 1+\eta(\sqrt{s}) \frac{C_4(\ycut)}{B_4(\ycut)}\:,
\eeq
are smaller for the Cambridge algorithm for small values of $\ycut$,
which is demonstrated in Fig.~3.

The smaller $K$ factors also mean smaller renormalization scheme
dependence, which can be seen from comparing Figs.~4 and 5.
The usual interpretation of the smaller scale dependence is that
the effect of the uncalculated higher orders are expected to be smaller
in the case of Cambridge clustering. It is interesting to note that in
the middle $\ycut$ region ($10^{-3.2}<\ycut<10^{-2}$), where the
hadronization corrections for the Cambridge clustering were found
significantly {\em larger} than for the Durham algorithm, the theoretical
uncertainty due to the renormalization scale ambiguity is
{\em smaller} for the Cambridge than that for the Durham clustering.
Of course, one has to keep in mind that the $\mu$-dependence bands are
not upper bounds on errors that arise from truncation of the
perturbation series, just suggestions.  In particular, if there is an
artificial narrowing of the $\mu$-dependence bands, e.g. at a crossover
point, they almost certainly do not represent the size of the
truncation error at that point.

Four-jet fractions decrease very rapidly with increasing resolution
parameter $\ycut$. As a result, most of the available four-jet data are
below $\ycut=0.01$.  It is well-known that for small values of $\ycut$
the fixed order perturbative prediction is not reliable, because the
expansion parameter $\as \ln^2 \ycut$ logarithmically enhances the
higher order corrections. One has to perform the all order resummation
of the leading and next-to-leading logarithmic (NLL) contributions.
This resummation is possible for the Durham algorithm using the coherent
branching formalism \cite{CDOTW} and the procedure is the same for the
Cambridge algorithm \cite{cambridge}.  The four-jet rate in the
next-to-leading logarithmic approximation is given by \cite{CDOTW}

\begin{figure}
\label{R4D}
\epsfxsize=8cm \epsfbox{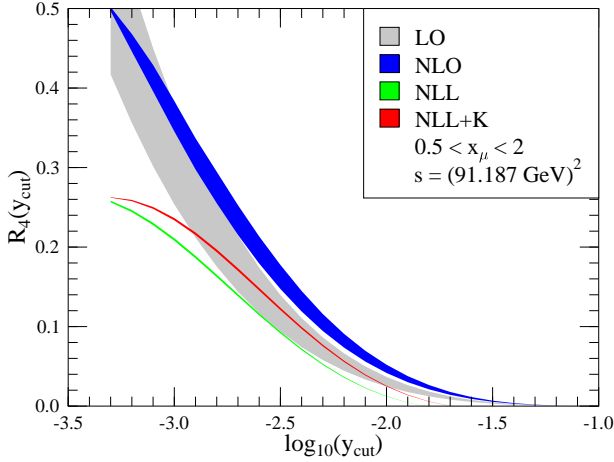}
\caption{The QCD prediction for the four-jet rate with Durham clustering
at Born level (light gray band) and at \NLO (dark band). The two narrow
bands show the four-jet rate in the NLL approximation ($K=0$, lower
band) and in improved NLL approximation (upper band) as explained in the
text. The bands indicate the theoretical uncertainty due to the
variation of the renormalization scale $x_\mu$ between 0.5 and 2.}
\end{figure}

\beeq
\label{R4NLL}
&&
R_4^{\rm NLL} = 2 [\Delta_q(Q)]^2
\Bigg[\left(\int_{Q_0}^Q\!\d q\,\Gamma_q(Q,q)\,\Delta_g(q,Q_0)\right)^2
\\ \nn && \quad
+\int_{Q_0}^Q\!\d q \,\Gamma_q(Q,q)\,\Delta_g(q,Q_0)
\\ \nn && \quad
\times \int_{Q_0}^q\!\d q'
\,\Big(\Gamma_g(q,q')\,\Delta_g(q',Q_0)
+\Gamma_f(q')\Delta_f(q',Q_0)\Big)\Bigg]\:.
\eeeq
In Eq.~(\ref{R4NLL}) the functions $\Delta_a(Q,Q_0)$ are the Sudakov form
factors which express the probability of parton branching evolution
from scale $Q_0=Q\sqrt{\ycut}$ to scale $Q$ without resolvable branching.
The Sudakov factors are defined in terms of the $P_{ab}(\as(q),z)$
vertex probabilities as follows
\beeq
\label{sudakov}
&&
\Delta_a(Q,Q_0) =
\\ \nn && \quad
\exp\left(-\sum_b \int_{Q_0}^Q\!\frac{\d q}{q}\int\!\d z\,
\frac{\as(q)}{2\pi}\,P_{ab}(\as(q),z)\right)\:.
\eeeq
It was shown in Ref.~\cite{DSansatz} that one can obtain an improved
theoretical prediction for the differential two-jet rate if the
vertex probabilities are taken at next-to-leading order
\cite{CMW}, which we also consider in our analysis:
\beeq
&&
P_{qq}(\as,z) =
C_F\left(\frac{1+z^2}{1-z}+\frac{\as}{2\pi}\,K\,\frac{2}{1-z}\right)\:,
\\ \nn &&
P_{gg}(\as,z) =
\\ \nn && \quad
2C_A\left(\frac{z}{1-z}+\frac{1-z}{z}+z(1-z)
+\frac{\as}{2\pi}\,K\,\frac{2}{z(1-z)}\right)\:,
\\ \nn &&
P_{gq}(\as,z) = T_R \Nf \left(z^2+(1-z)^2\right)\:.
\eeeq
\begin{figure}
\label{R4C}
\epsfxsize=8cm \epsfbox{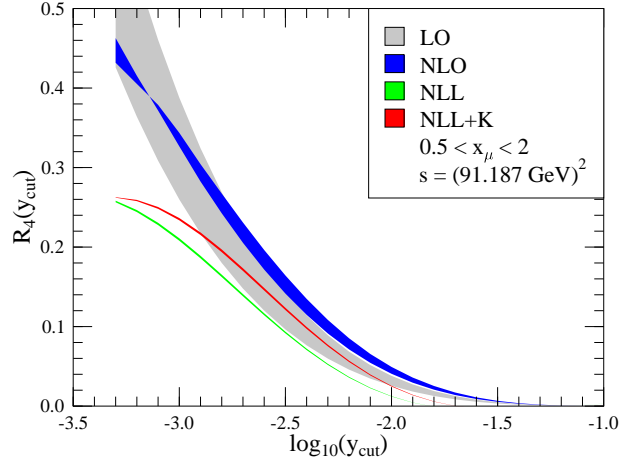}
\caption{The QCD prediction for the four-jet rate with Cambridge clustering
at Born level (light gray band) and at \NLO (dark band). The two narrow
bands show the four-jet rate in the NLL approximation ($K=0$, lower
band) and in improved NLL approximation (upper band) as explained in the
text. The bands indicate the theoretical uncertainty due to the
variation of the renormalization scale $x_\mu$ between 0.5 and 2.}
\end{figure}

The $K$ coefficient is renormalization scheme dependent. In the
$\overline{{\rm MS}}$ scheme it is given by \cite{Kcoeff}
\beq
K=C_A\left(\frac{67}{18}-\frac{\pi^2}{6}\right)-\frac{10}{9}T_R\Nf\:.
\eeq
Performing the $z$ integral in Eq.~(\ref{sudakov}), one obtains the
Sudakov factors as integrals of the emission probabilities
$\Gamma_a(Q,q)$ in the following form:
\beeq
&&
\Delta_q(Q,Q_0)=\exp\left(-\int_{Q_0}^Q\!\d q\,\Gamma_q(Q,q)\right)\:,
\\ &&
\Delta_g(Q,Q_0)=
\\ \nn && \qquad
\exp\left(-\int_{Q_0}^Q\!\d q\,\left[\Gamma_g(Q,q)+\Gamma_f(q)\right]\right)\:,
\\ &&
\Delta_f(Q,Q_0)= \frac{\left[\Delta_q(Q,Q_0)\right]^2}{\Delta_g(Q,Q_0)}\:,
\eeeq
and the NLL emission probabilities are 
\beeq
&&
\Gamma_q(Q,q) =
\\ \nn && \qquad
\frac{2 C_F}{\pi}\frac{\as(q)}{q}
\left[\left(1+\frac{\as(q)}{2\pi}\,K\right)\ln\frac{Q}{q}
-\frac{3}{4}\right]\:,
\\ &&
\Gamma_g(Q,q) =
\\ \nn && \qquad
\frac{2 C_A}{\pi}\frac{\as(q)}{q}
\left[\left(1+\frac{\as(q)}{2\pi}\,K\right)\ln\frac{Q}{q}
-\frac{11}{12}\right]\:,
\\ &&
\Gamma_f(Q,q) =
\frac{\Nf}{3\pi}\frac{\as(q)}{q}\:.
\eeeq
We relate the $\as(q)$ strong coupling appearing in the emission
probabilities to the strong coupling at the relevant renormalization
scale, $\as(\mu)$, according to the one-loop formula
\beq
\as(q) = \frac{\as(\mu)}{w(q,\mu)}\:,
\eeq
where $w(q,q_0)$ was defined in Eq.~(\ref{wqq0}),
and we use Eq.~(\ref{twoloopas}) for expressing $\as(\mu)$ in terms of
$\as(M_Z)=0.118$. We could also use a two-loop formula for $\as(q)$,
but the result would differ only in subleading logarithms.

The result of this resummation together with its renormalization scale
dependence is also shown in Figs.~4 and 5 (narrow bands). The lower band
corresponds to the usual NLL approximation ($K=0$), and the upper band
is the result of the improved resummation. We can see clearly from the
figures that the fixed-order and the NLL approximations differ
significantly. One expects that for large values of $\ycut$ the former,
and for small values of $\ycut$ the latter is the reliable description,
therefore, the two results have to matched.

The Durham and Cambridge four-jet rates can be resummed at leading and
next-to-leading logarithmic order, but they do not satisfy a simple
exponentiation \cite{Catani}. For observables that do not exponentiate
the viable matching schemes are the R matching or the modified R
matching \cite{CDOTW,ALEPH}. We use R matching according to the following
formula:
\beeq
&&
R_4^{\rm R-match} = R_4^{\rm NLL} +
%\\ \nn && \qquad
\Bigg[\eta^2
\left(B_4 - B_4^{\rm NLL}\right)
\\ \nn && \qquad\qquad\qquad
+\eta^3
\left(C_4 - C_4^{\rm NLL}
-\frac32\left(B_4 - B_4^{\rm NLL}\right)\right)\Bigg]\:,
\eeeq
where $B_4^{\rm NLL}$ and $C_4^{\rm NLL}$ are the
%O($\alpha_s^2$) and O($\alpha_s^3$)
coefficients in the expansion of $R_4^{\rm NLL}$
as in Eq.~(\ref{R4}).

In Fig.~6 we show the theoretical prediction at the various levels of
approximation: in fixed order perturbation theory at Born level (LO),
at \NLO (NLO), resummed and R-matched prediction (NLO+NLL) and improved
resummed and R-matched prediction (NLO+NLL+K).  Also shown is the
four-jet rate measured by the ALEPH collaboration at the $Z^0$ peak
\cite{ALEPHR4} corrected to parton level using the PYTHIA Monte Carlo
\cite{PYTHIA}. We used bin-by-bin correction and the consistency of the
correction was checked by using the HERWIG Monte Carlo \cite{HERWIG}.
The two programs gave the same correction factor within statistical
error. The errors of the data are the scaled errors of the published
hadron level data, and we did not include any systematic error due to
the hadron to parton correction. In the inset we indicated the
renormalization scale dependence of the `NLO+NLL+K' prediction.

\begin{figure}
\epsfxsize=8cm \epsfbox{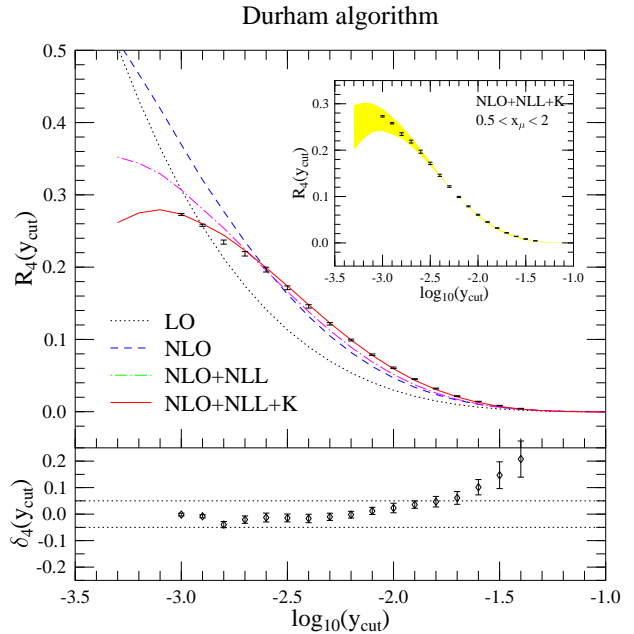}
\caption{The QCD prediction for the four-jet rate with Durham clustering
in fixed order perturbation theory at leading (dotted) and \NLO
(dashed), and fixed order matched with resummed (dashed-dotted) and
improved resummed (solid) calculation compared to ALEPH data obtained at
the $Z^0$ peak and corrected to parton level (errorbars). The
renormalization scale is set to $x_\mu=1$. The lower plot shows the
relative difference $\delta_4=$ (data--theory)/theory, where theory means
the \NLO prediction matched with improved resummed calculation at
$x_\mu=1$. The inset shows the renormalization scale dependence of the
`theory' prediction with scale variation $0.5<x_\mu<2$.}
\end{figure}

Fig.~6 deserves several remarks. First of all, we see that the inclusion
of the radiative corrections improves the fixed order description of the
data using the natural scale $x_\mu=1$ for larger values of $\ycut$.
Secondly, the importance of resummation in the small $\ycut$ region is
clearly seen, but it is still not sufficient to describe the data at the
natural scale, neglected subleading terms are still important.\footnote{
Our `NLO+NLL' results differ from those in Ref.~\cite{DSjets}, where
$\as(q)$ in calculating $R_4^{\rm NLL}$ was kept at the fix $\as(M_Z)$
value \cite{DSpriv}.} On the other hand, the improved resummation seems
to take into account just the right amount of subleading terms and it
makes the agreement between data and theory almost perfect over the whole
$\ycut$ region as can be seen from the lower plot. (Although for $\ycut
> 10^{-1.7}$ $\delta_4$ falls outside the $\pm 5\,\%$ band, one should
keep in mind that in this region the error of the hadron to parton
correction is very large. Also, for the `NLO+NLL+K' prediction we found
remarkably small scale dependence for $\ycut > 10^{-3}$. This feature,
however, should be taken with care. The improvement, obtained by
including the two-loop coefficient K, affects NNLL terms, but there are
other contributions of the same order that are not taken into account
(e.g., \NLO running of $\as$ and other dynamical effects), which is not
the case for the 2-jet rate. The scale dependence of the `NLO+NLL+K'
result would consistently be under control only after the inclusion of
the complete set of NNLL terms.

Finally it is worth noting that for $\ycut=10^{-2.6}$  both PYTHIA and
HERWIG yield less than 2\,\% hadronization correction. At the same
value of the resolution parameter the theoretical prediction is
insensitive to corrections beyond next-to-leading order (the NLO,
NLO+NLL, NLO+NLL+K curves cross, the renormalization scale dependence
is small), therefore, at this accidental value of $\ycut$ the \NLO
prediction agrees perfectly with the hadron level data.

\subsection{Four-jet event shapes}

Four-jet event shapes were used extensively by the LEP collaborations for
QCD studies \cite{ALEPHR4,LEPshapes}.
In this subsection we consider four shape variables, the $\yflip$
distributions for the Durham and Cambridge algorithms, the thrust minor
($T_{\rm min}$) and the C parameter for C values above 0.75, which are
often used in the experimental analyses.

In the case of event shape distributions we multiply the normalized
cross section with the value of the event shape parameter, so we use
the parametrisation
\beeq
\label{sigmaO4}
&&\Sigma(O_4)\equiv O_4 \frac{1}{\sigma_0}\frac{\d \sigma}{\d O_4}(O_4)
\\ \nn &&\qquad
= \eta(\mu)^2 B(O_4)
+ \eta(\mu)^3\left[
B(O_4)\frac{\beta_0}{C_F} \ln x_\mu^2 + C(O_4)\right]
\eeeq
instead of Eq.~(\ref{nloxsec2}), in which case the average value of the
shape variable is easily obtained from the differential distribution:
\beq
<O_4>_\delta = \int_\delta^1\!\d O_4\,\Sigma(O_4)\:.
\eeq
Using this parametrisation we define the $K$ factors of the
differential distribution as
\beq
K(O_4) = 1+\eta(\sqrt{s})\,\frac{C(O_4)}{B(O_4)}\:.
\eeq
In the following we plot the physical cross sections $\Sigma(O_4)$,
the $K(O_4)$ factors and tabulate the correction functions $C(O_4)$ for
$O_4 = \yflip$, $T_{\rm min}$ and C.

The $\yflip$ value denotes the transition value for $\ycut$ at which,
when decreasing $\ycut$, the classification of a given event changes
from three jets to four jets. The advantage of this variable over the
differential four-jet rate is that this variable can be defined on an
event by event basis. Depending on the actual resolution variable one
obtains the $\yflipD$ distribution for the Durham clustering and the
$\yflipC$ distribution for the Cambridge clustering. We calculated the
$B(\yflip)$ and $C(\yflip)$ functions for both algorithms. The
$B(\yflip)$ values equal the $\ycut B(\ycut)$ values when $\yflip=\ycut$,
therefore, we tabulate only the $C(\yflip)$ functions for the two
algorithms in Table~II. 

\vbox{\begin{table}             
\caption{Correction functions to the differential distributions of the
$\yflip$ variables for the Durham and Cambridge algorithm. The parameter
values are at the lower edge of the corresponding histogram bin.}
\begin{tabular}{ccc}
  $y_4$ & $C_{y_4}^{\rm D}$ & $C_{y_4}^{\rm C}$ \\
\tableline                        
0.000 & \res{2.523}{0.425}{3} & \res{1.064}{0.350}{3} \\
0.005 & \res{2.212}{0.017}{3} & \res{1.857}{0.019}{3} \\
0.010 & \res{1.376}{0.009}{3} & \res{1.166}{0.011}{3} \\
0.015 & \res{9.429}{0.071}{2} & \res{8.144}{0.080}{2} \\
0.020 & \res{6.799}{0.070}{2} & \res{5.855}{0.062}{2} \\
0.025 & \res{4.930}{0.063}{2} & \res{4.346}{0.051}{2} \\
0.030 & \res{3.760}{0.042}{2} & \res{3.293}{0.043}{2} \\
0.035 & \res{2.885}{0.037}{2} & \res{2.553}{0.039}{2} \\
0.040 & \res{2.164}{0.033}{2} & \res{1.947}{0.034}{2} \\
0.045 & \res{1.754}{0.026}{2} & \res{1.580}{0.027}{2} \\
0.050 & \res{1.314}{0.025}{2} & \res{1.202}{0.025}{2} \\
0.055 & \res{1.024}{0.021}{2} & \res{9.508}{0.213}{1} \\
0.060 & \res{8.293}{0.292}{1} & \res{7.692}{0.296}{1} \\
0.065 & \res{6.307}{0.300}{1} & \res{5.945}{0.304}{1} \\
0.070 & \res{4.636}{0.180}{1} & \res{4.445}{0.184}{1} \\
0.075 & \res{3.516}{0.117}{1} & \res{3.430}{0.114}{1} \\
0.080 & \res{2.673}{0.115}{1} & \res{2.560}{0.110}{1} \\
0.085 & \res{2.271}{0.216}{1} & \res{2.214}{0.217}{1} \\
0.090 & \res{1.412}{0.204}{1} & \res{1.395}{0.206}{1} \\
0.095 & \res{1.085}{0.057}{1} & \res{1.056}{0.059}{1} \\
0.100 & \res{7.412}{0.584}{0} & \res{7.377}{0.588}{0} \\
0.105 & \res{5.069}{0.537}{0} & \res{5.121}{0.529}{0} \\
0.110 & \res{2.817}{0.400}{0} & \res{2.914}{0.399}{0} \\
0.115 & \res{2.652}{0.329}{0} & \res{2.428}{0.301}{0} \\
0.120 & \res{1.353}{0.221}{0} & \res{1.516}{0.183}{0} \\
\end{tabular} 
\end{table}
\noindent

 }

We show the \NLO perturbative prediction in QCD for $\Sigma(\yflip)$ in
Fig.~7.  In the same figure, the inset shows the $K(\yflip)$ factors of
the distributions. The physical cross sections for the two algorithms
are very similar. The $K(\yflip)$ factors are quite large, but much
smaller than in the case of other four-jet event shape distributions.
They depend weakly on the $\yflip$ value for $\yflip>0.1$ and decrease
rapidly with decreasing $\yflip$ below $\yflip=0.1$. In the case of the
Cambridge algorithm the radiative corrections are 15--30\,\% smaller
than those for the Durham algorithm.

In order to define $T_{\rm min}$, we first have to define the thrust and
thrust major axes \cite{Tmin}. The thrust axis $\vec{n}_T$ is the
direction $\vec{n}$ which maximizes the expression
\beq
\label{thrust}
T=\max_{\vec{n}_T}
\left(\frac{\sum_a |\vec{p}_a\cdot\vec{n}|}{\sum_a |\vec{p}_a|}\right),
\eeq
\begin{figure}
\label{yflip}
\epsfxsize=8cm \epsfbox{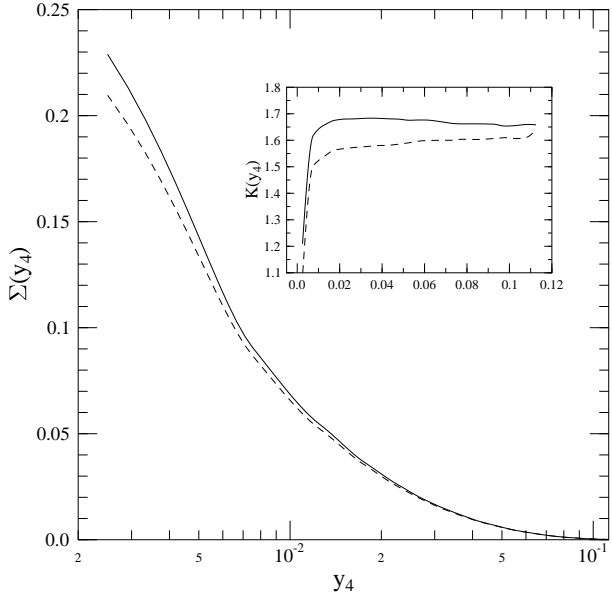}
\caption{The \NLO QCD prediction for the $\yflipD$ (solid) and $\yflipC$
(dashed) differential distributions with renormalization scale $x_\mu=1$.
The inset shows the $K$ factors of the distributions.}
\end{figure}
\noindent
where the sum runs over all final state hadrons (or partons). The thrust
major axis is a three-vector $\vec{n}_{T_{\rm maj}}$ for which the
expression in Eq.~(\ref{thrust}) is maximal with the constraint that
$\vec{n}_{T_{\rm maj}}$ is perpendicular to $\vec{n}_T$, $\vec{n}_{T_{\rm
maj}}\cdot\vec{n}_T=0$. In order to obtain the value of $T_{\rm min}$, one
evaluates the expression in the parentheses for a vector perpendicular
to both $\vec{n}_{T_{\rm maj}}$ and $\vec{n}_T$.

The C parameter \cite{Dpar} is derived from the eigenvalues of the
infrared safe momentum tensor                               
\beq
\theta^{ij} = 
\sum_a \frac{p_a^i p_a^j}{|\vec{p}_a|}\bigg/\sum_a |\vec{p}_a|,
\eeq 
where the sum on $a$ runs over all final state hadrons and $p_a^i$ is the
$i$th component of the three-momentum $\vec{p}_a$ of hadron $a$ in the
c.m. system. The tensor $\theta$ is normalized to have unit trace. In
terms of the eigenvalues $\lambda_i$ of the $3 \times 3$ matrix
$\theta$, the global shape parameter C is defined as
\beq\label{Ddef}
{\rm C} = 
3\,(\lambda_1 \lambda_2 + \lambda_2 \lambda_3 + \lambda_3 \lambda_1)\:.
\eeq
The kinematical limit of the C parameter for three-parton processes is 
C = 0.75. Therefore, in the region C $\in [0.75,1]$ the four-parton
processes contribute to the leading order prediction, and our program
is capable to calculate the radiative correction to the distribution.
The results of such a calculation for the Born functions $B_{T_{\rm min}}$
and $B_{\rm C}$ agree with the known results (see e.g., \cite{KNMW})
The $C_{T_{\rm min}}$ and $C_{\rm C}$ correction functions are given in
Table~III.

\vbox{\begin{table}             
\caption{Correction functions to the differential distributions of the 
$T_{\rm min}$ and C parameter event shape variables. The parameter 
values are at the lower edge of the corresponding histogram bin.}
\begin{tabular}{cccc}
 $T_{\rm min}$ & $C_{T_{\rm min}}$ & ${\rm C}$ & $C_{\rm C}$ \\
\tableline                        
0.00 &      --               &  0.75 & \res{4.775}{1.100}{3} \\
0.02 & \res{3.319}{0.270}{4} &  0.76 & \res{6.082}{0.160}{3} \\
0.04 & \res{2.381}{0.082}{4} &  0.77 & \res{4.610}{0.089}{3} \\
0.06 & \res{1.652}{0.038}{4} &  0.78 & \res{3.663}{0.063}{3} \\
0.08 & \res{1.172}{0.025}{4} &  0.79 & \res{2.904}{0.042}{3} \\
0.10 & \res{8.600}{0.130}{3} &  0.80 & \res{2.406}{0.031}{3} \\
0.12 & \res{6.488}{0.100}{3} &  0.81 & \res{1.948}{0.026}{3} \\
0.14 & \res{4.695}{0.077}{3} &  0.82 & \res{1.625}{0.024}{3} \\
0.16 & \res{3.499}{0.042}{3} &  0.83 & \res{1.365}{0.023}{3} \\
0.18 & \res{2.684}{0.027}{3} &  0.84 & \res{1.135}{0.019}{3} \\
0.20 & \res{2.010}{0.021}{3} &  0.85 & \res{9.194}{0.130}{2} \\
0.22 & \res{1.498}{0.017}{3} &  0.86 & \res{7.906}{0.110}{2} \\
0.24 & \res{1.122}{0.013}{3} &  0.87 & \res{6.293}{0.092}{2} \\
0.26 & \res{8.247}{0.100}{2} &  0.88 & \res{5.217}{0.084}{2} \\
0.28 & \res{6.093}{0.074}{2} &  0.89 & \res{4.296}{0.066}{2} \\
0.30 & \res{4.501}{0.180}{2} &  0.90 & \res{3.391}{0.052}{2} \\
0.32 & \res{3.026}{0.057}{2} &  0.91 & \res{2.815}{0.061}{2} \\
0.34 & \res{2.229}{0.050}{2} &  0.92 & \res{2.075}{0.057}{2} \\
0.36 & \res{1.549}{0.046}{2} &  0.93 & \res{1.626}{0.032}{2} \\
0.38 & \res{1.095}{0.028}{2} &  0.94 & \res{1.221}{0.026}{2} \\
0.40 & \res{7.100}{0.210}{1} &  0.95 & \res{8.154}{0.260}{1} \\
0.42 & \res{4.437}{0.180}{1} &  0.96 & \res{5.193}{0.190}{1} \\
0.44 & \res{2.684}{0.190}{1} &  0.97 & \res{3.165}{0.130}{1} \\
0.46 & \res{1.439}{0.150}{1} &  0.98 & \res{1.312}{0.094}{1} \\
0.48 & \res{6.447}{0.560}{0} &  0.99 & \res{2.769}{0.260}{0} \\
\end{tabular}                                               
\end{table}
\noindent

 }

In the case of event shape differential distributions the \NLO
corrections should logarithmically diverge at the edge of the phase
space. This divergence occurs at zero for the $y_4$ and $T_{\rm min}$
distributions and is regularized by the multiplication with the value of
the variable (see Eq.~(\ref{sigmaO4})). This is not the case for the C
parameter, because it diverges at C = 0.75. Nevertheless, we obtained a
finite and positive contribution in the first bin owing to bin smearing
as we have checked explicitly by refining the bin width.

Figs.~8 and 9 show the leading and \NLO QCD prediction for the
$T_{\rm min}$ and C parameter differential distributions at $x_\mu=1$. 
The insets show the $K$ factors which are large in both cases
indicating 100\,\% or larger radiative corrections. As a result, the
renormalization scale dependence remains large, only the absolute
normalization of the distributions increases with a factor of more than
2 with the inclusion of the radiative corrections. This feature is
demonstrated in Fig.~10, where
we show the scale dependence of the
leading and \NLO prediction for the average value of the thrust minor
(above $T_{\rm min} = 0.02$) and C parameter (above C = 0.75). The
leading and \NLO curves run almost parallel down to $x_\mu\simeq 0.2$,
only the latter is shifted to larger values.

\begin{figure}
\label{Tmin}
\epsfxsize=8cm \epsfbox{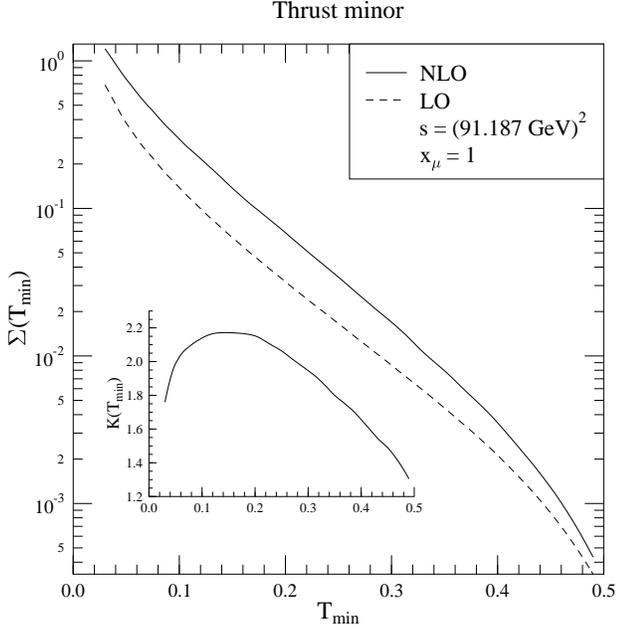}
\caption{The leading-order (dashed) and the \NLO (solid) QCD prediction
for the $T_{\rm min}$ variable with renormalization scale $x_\mu=1$. 
The inset shows the $K$ factor of the distribution.}
\end{figure}
\begin{figure}
\label{Cpar}
\epsfxsize=8cm \epsfbox{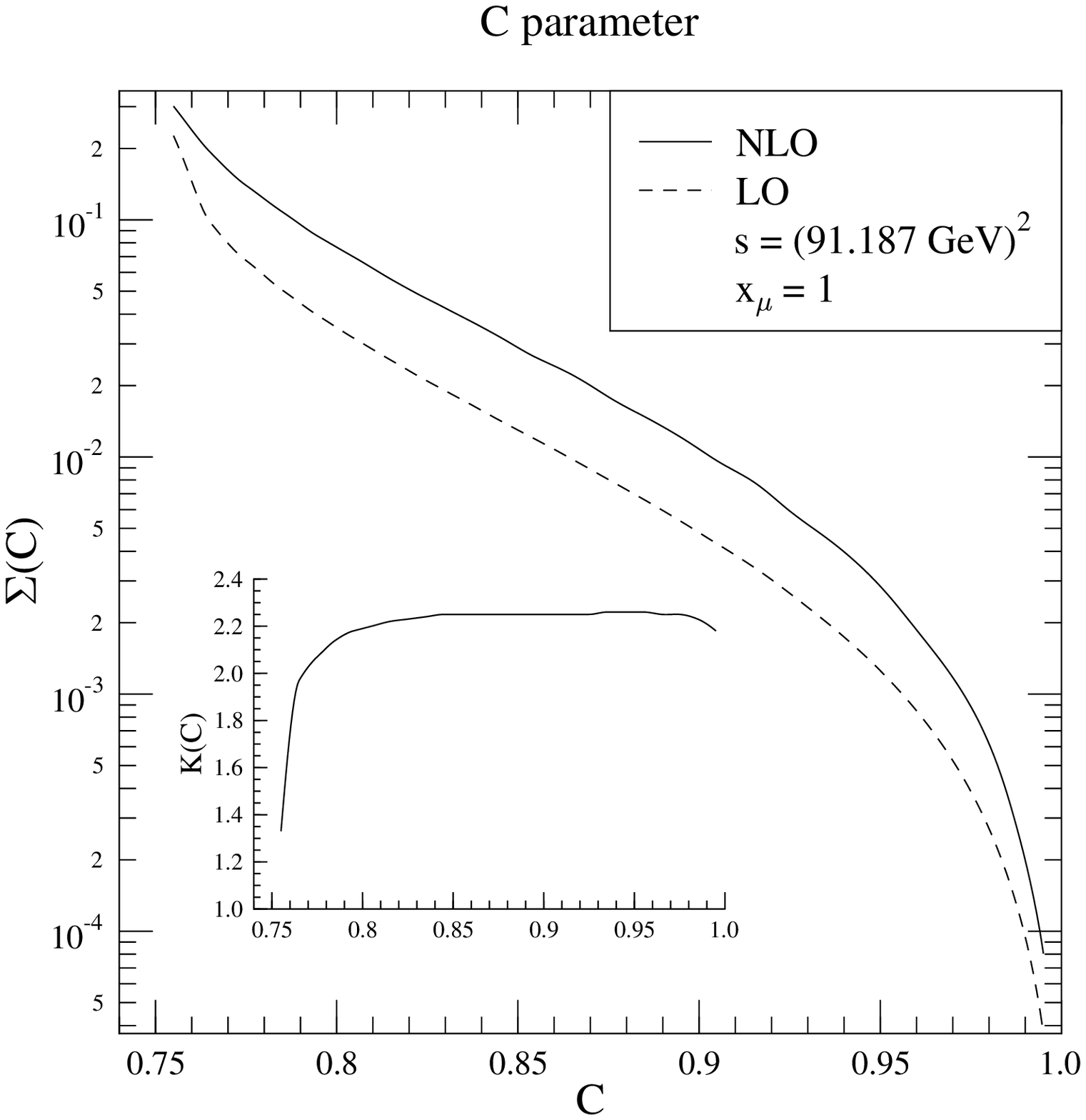}
\caption{The leading-order (dashed) and the \NLO (solid) QCD prediction
for the C parameter with renormalization scale $x_\mu=1$.  The inset
shows the $K$ factor of the distribution.}
\end{figure}

\begin{figure}
\label{TCmudep}
\epsfxsize=8cm \epsfbox{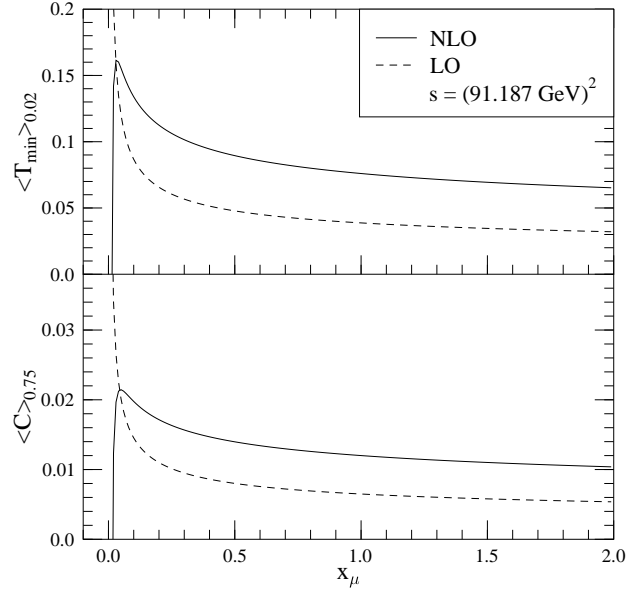}
\caption{The renormalization scale dependence of the average values of
$<T_{\rm min}>_{0.02}$ and $<{\rm C}>_{0.75}$ at leading and next-to-leading
order.}
\end{figure}

\subsection{Radiative corrections to four-jet observables: summary}

In this subsection we summarize the results of our radiative correction
calculations for the various four-jet like distributions presented in
previous publications and in this article.

The QCD prediction at tree level (with renormalization scale $x_\mu=1$)
is in general falls significantly below the measured values for
unnormalised distributions of four-jet observables. Consequently, the
calculation of the \NLO corrections to
these cross sections is indispensable for attempting a serious comparison
between data and theory. Our calculations show that the corrections are
very large and the agreement in the comparison improves considerably with
the inclusion of the radiative corrections. In particular, we found the
following features:\\
(1)
In the case of four-jet rates, the radiative corrections are about
100\,\% for JADE-type clustering algorithms \cite{DSjets,NTDpar},
while for the Durham algorithm it is less than 60\,\% and even smaller
for the Cambridge algorithm. The scale dependence for the latter
algorithms is substantially reduced. The agreement between data and
theory for the Durham clustering is very good and extends to small
values of $\ycut$ when one matches the fixed order prediction with
improved resummed next-to-leading logarithmic approximation. \\
(2)
In the case of event shape variables the corrections are usually more
than 100\,\% (the $K$ factors are larger than 2).  The residual
renormalization scale dependence is large indicating that even higher
orders are important. One may conclude that, with the exception of the
jet-related $\yflip$ distributions, these distributions cannot be
reliably calculated in fixed order perturbation theory and cannot be used
for precision tests of QCD.\\
(3)
In the case of normalized angular distributions the corrections are small
as expected (the $K$ factors are close to 1) \cite{Signer,NTangulars}.
The renormalization scale dependence is small, which however, does not
mean that the effect of the radiative corrections on the measurement of
the QCD color charges is negligible. According to Ref.~\cite{NTangulars},
the measured value of the $T_R/C_F$ ratio may differ up to 25\,\% when
leading, or \NLO QCD predictions are used in the color charge fits.

\section{Conclusions}

This paper dealt with the \NLO calculation of four-jet observables
in electron-positron annihilation. We gave details of the
analytical calculation that lead to the construction of a Monte Carlo
program \cite{debrecen} which can be used to calculate the differential
distribution of any four-jet observable at the O($\alpha_s^3$)
accuracy. The dipole method was used for achieving the analytical
cancellation of infrared divergences. We described that modification of
the algorithm which we found useful from numerical point of view.
However, the modification is not essential as far as the theory is
concerned.

Compact formulas were presented for the Born-level five-parton helicity
amplitudes and for the Born-level four-parton color-correlated
matrix elements which are necessary for other \NLO calculations, such
as the \NLO cross section of three-jet production in deep inelastic
scattering and that of vector boson plus two-jet production in hadron
collisions. We also gave a group independent decomposition of the
Born-level five-parton matrix elements.

We calculated the \NLO corrections to the four-jet rates with the
Durham and Cambridge jet clustering algorithms and to the differential
distributions of the $\yflip$, thrust minor and C parameter (at C $\ge$
0.75) four-jet shape variables. In the case of four-jet rates the
radiative corrections were found to be large, but just acceptable. The 
renormalization scale dependence decreased significantly and the fixed
order result matched with the next-to-leading logarithmic approximation
gave remarkably good agreement with LEP data over a wide range of the
resolution variable. The high level of agreement implies that the QCD
four-jet background to $W^\pm$ pair production at higher center of mass
energies can be predicted in perturbation theory reliably. In the case
of event shape variables the radiative corrections and the
renormalization scale dependence are unacceptably large suggesting that
the \NLO prediction is not reliable and even higher orders are important.

\acknowledgments

We thank S. Catani for comments on the manuscript.
This work was supported in part by the EU Fourth Framework Programme
``Training and Mobility of Researchers'', Network ``Quantum
Chromodynamics and the Deep Structure of Elementary Particles'',
contract FMRX-CT98-0194 (DG 12 - MIHT), as well as by the Hungarian
Scientific Research Fund grant OTKA T-025482, the Research Group in
Physics of the Hungarian Academy of Sciences, Debrecen and the
Universitas Foundation of the Hungarian Commercial Bank.

\appendix
\section{Helicity amplitudes}

In this appendix we present analytic formulas for the four- and
five-parton tree-level helicity amplitudes of the relevant
subprocesses. These amplitudes were first calculated in
Refs.~\cite{a5parton}. The reason for presenting our results here is
twofold.  On one hand we express the relevant color subamplitudes in
terms of Weyl spinors $\ket{k\pm}$, which were also employed in the
case of the one-loop four-parton amplitudes \cite{BDK2q2g}, while on
the other we found that our expressions in the case of the four-quark
processes are more compact and the corresponding computer code is
faster than earlier ones. Another new feature of the
amplitudes in this appendix is that we allow for the existence of light
fermionic degrees of freedom in the adjoint representation of the gauge
group (light gluinos).  In calculating the amplitudes, we used quark
and gluon currents \cite{BGcurrents,MP} and standard helicity
techniques \cite{helicity,MP}.

We consider three subprocesses, each involving a vector boson $V(Q)$
carrying total four-momentum $Q$ and $n$ QCD partons ($n=4$, or 5
here). The first subprocess is the production of a quark-antiquark pair
and $n-2$ gluons. The second one is the production of two
quark-antiquark pairs (of equal, or unequal flavor) and $n-4$ gluons.
Finally, the third process is the production of a quark-antiquark pair,
a light-gluino pair and $n-4$ gluons:
\beeq
&&
\label{Vqq}
\ell^+(-p_\ell)+\ell^-(-p_{\bar \ell}) \to V(Q) \to
\\ \nn && \qquad
q(p_1)+g_1(p_2)+\dots+g_{n-2}(p_{n-1})+\bar{q}(p_n)\:,
\vspaceinarray
\\ &&
\label{VqqQQ}
\ell^+(-p_\ell)+\ell^-(-p_{\bar \ell}) \to V(Q) \to
\\ \nn && \qquad
q(p_1)+\bar{q}(p_2)+Q(p_3)+\bar{Q}(p_4)
\\ \nn && \qquad
+g_1(p_5)+\dots+g_{n-4}(p_n)\:,
\vspaceinarray
\\ &&
\label{Vqqgtgt}
\ell^+(-p_\ell)+\ell^-(-p_{\bar \ell}) \to V(Q) \to
\\ \nn && \qquad
q(p_1)+\bar{q}(p_2)+\gt(p_3)+\gt(p_4)
\\ \nn && \qquad
+g_1(p_5)+\dots+g_{n-4}(p_n)\:.
\eeeq
We have chosen the crossing invariant all particle outgoing kinematics
with corresponding particle-antiparticle assignment,
therefore, momentum conservation means
\beq                                                              
p_\ell+p_{\bar{\ell}}+p_1+p_2+p_3+p_4+p_5+\ldots+p_n=0\:.
\eeq

We shall express the amplitudes in terms of color subamplitudes. In
the case of process (\ref{Vqq}), the color basis is chosen to be product
of generators in the fundamental representation (in this appendix we use
the normalization $T_R=1$ in ${\rm Tr}(t^a t^b)=T_R\,\delta^{ab}$ for the
generators of the symmetry group), therefore, the helicity amplitudes
have the decomposition:
\beeq
\label{a2q}
&&
\nket{1_f^{h_1},2_g^{h_2},\ldots,n_{\bar f}^{h_n}} =
\\ \nn&& \qquad
\sum_{\{2,\dots,n-1\}}
\left(t^{a_2}\cdot\dots\cdot t^{a_{n-1}}\right)_{i_1{\bar i}_n}\,
m(1_f^{h_1},\dots,n_{\bar f}^{h_n})\:,
\eeeq
where $\{2,\dots,n-1\}$ denotes all permutations of the labels
$(2,\dots,n-1)$ and $m(1,\dots,n)$ are the color subamplitudes.
In Eq.~(\ref{a2q}) and in the following formulas the lepton labels are
suppressed.

In the case of the four-fermion subprocesses (processes (\ref{VqqQQ}) and
(\ref{Vqqgtgt})) we decompose the helicity amplitudes as follows:
\beeq
&&
\nket{1_{f_1}^{h_1},2_{f_2}^{h_2},3_{f_3}^{h_3},4_{f_4}^{h_4},
5_g^{h_5},\dots,n_g^{h_n}}=
\\ \nn &&\qquad
\sum_{\{5,\dots,n\}}
\sum_{\{1,3\}}(-1)^P
\sum_{\{2,4\}}(-1)^P
{\cal A}_{n}(1,2,3,4,5,\dots,n)\:.
\eeeq
where $P=0$ if the elements are in the canonical order ((1,3), or (2,4))
and $P=1$ if the elements are permuted ((3,1), or (4,2)).  The partial
amplitudes ${\cal A}_n$ can be decomposed further in color space. In
the case of four-quark production,
\beeq
&&
{\cal A}_4(1_q,2_\qb,3_Q,4_\Qb) =
\\ \nn && \qquad
T(1,2,3,4)\,M(1_{f_1}^{h_1},2_{f_2}^{h_2},3_{f_3}^{h_3},4_{f_4}^{h_4})\:,
\eeeq
where $M(1_{f_1}^{h_1},2_{f_2}^{h_2},3_{f_3}^{h_3},4_{f_4}^{h_4})$ are
the color subamplitudes and $T(1,2,3,4)$ is defined by
\beeq
T(1,2,3,4) &=& \sum_{b=1}^{\NA}t^b_{i_1{\bar i}_2}\,t^b_{i_3{\bar i}_4}\:.
\eeeq
In the case of four-quark plus one-gluon production, there are four
independent basis vectors in color space:
\beeq
&&
{\cal A}_5(1_q,2_\qb,3_Q,4_\Qb,5_g) =
\\ \nn && \qquad
\sum_{i=1}^4 T_i(1,2,3,4,5)\,
M_i(1_{f_1}^{h_1},2_{f_2}^{h_2},3_{f_3}^{h_3},4_{f_4}^{h_4},5^{h_5})\:,
\eeeq
where $T_i(1,2,3,4,5)$ are the color basis vectors:
\beeq
&&
T_1(1,2,3,4,5) =
\sum_{b=1}^{\NA}(t^{a_5}t^b)_{i_1{\bar i}_2}\,t^b_{i_3{\bar i}_4}\:,
%\vspaceinarray
\\ &&
T_2(1,2,3,4,5) =
\sum_{b=1}^{\NA}(t^bt^{a_5})_{i_1{\bar i}_2}\,t^b_{i_3{\bar i}_4}\:,
%\vspaceinarray
\\ &&
T_3(1,2,3,4,5) =
\sum_{b=1}^{\NA}t^b_{i_1{\bar i}_2}\,(t^{a_5}t^b)_{i_3{\bar i}_4}\:,
%\vspaceinarray
\\ &&
T_4(1,2,3,4,5) =
\sum_{b=1}^{\NA}t^b_{i_1{\bar i}_2}\,(t^bt^{a_5})_{i_3{\bar i}_4}\:.
\eeeq

The partial amplitudes for the process (\ref{Vqqgtgt}) can be written in
terms of the color subamplitudes of the process (\ref{VqqQQ}), only the
color basis differs. When $n=4$,
\beeq
&&
{\cal A}_4(1_q, 2_\qb, 3_\gt, 4_\gt) =
\\ \nn && \qquad
\widetilde{T}(1,2,3,4)\,
\widetilde{M}(1_{f_1}^{h_1},2_{f_2}^{h_2},3^{h_3},4^{h_4})\:,
\eeeq
where
\beq
\widetilde{T}(1,2,3,4) =
\sum_{b=1}^{\NA}t^b_{i_1{\bar i}_2}\, F^b_{a_3a_4}\:.
\eeq
Finally, for $n=5$ we have
\beeq
&&
{\cal A}_5(1_q, 2_{\bar{q}}, 3_{\tilde{g}}, 4_{\tilde{g}},5_g) =               
\\ \nn && \qquad
\sum_{i=1}^4 \widetilde{T}_i(1,2,3,4,5)\,
\widetilde{M}_i(1_{f_1}^{h_1},2_{f_2}^{h_2},3^{h_3},4^{h_4},5^{h_5})\:,
\eeeq
where
\beeq
&&
\widetilde{T}_1(1,2,3,4,5) =
\sum_{b=1}^{\NA}(t^{a_5}t^b)_{i_1{\bar i}_2}\,F^b_{a_3a_4}\:,
%\vspaceinarray
\\ &&
\widetilde{T}_2(1,2,3,4,5) =
\sum_{b=1}^{\NA}(t^bt^{a_5})_{i_1{\bar i}_2}\,F^b_{a_3a_4}\:,
%\vspaceinarray
\\ &&
\widetilde{T}_3(1,2,3,4,5) =
\sum_{b=1}^{\NA}t^b_{i_1{\bar i}_2}\,(F^{a_5}F^b)_{a_3a_4}\:,
%\vspaceinarray
\\ &&
\widetilde{T}_4(1,2,3,4,5) =
\sum_{b=1}^{\NA}t^b_{i_1{\bar i}_2}\,(F^bF^{a_5})_{a_3a_4}\:.
\eeeq

In the following subsections we give explicit formulas for the color
subamplitudes with a common coefficient factored out:
\beeq
&&
m(1_{f_1}^{h_1},\dots,n_{f_n}^{h_n})=
\\ \nn && \qquad
2\,e^2\,g^{(n-2)}\,C_{f_1f_n}^{h_\ell, h_1}\,\frac{\i}{s}\,
A(1^{h_1},\dots,n^{h_n}) \:,
\vspaceinarray\\ &&
M(1_{f_1}^{h_1},2_{f_2}^{-h_1},3_{f_3}^{h_3},4_{f_4}^{-h_4}) =
\\ \nn && \qquad
2\,e^2\,g^2\,C_{f_1f_2}^{h_\ell, h_1}\,\delta_{f_3f_4}\,\frac{\i}{s}\,
A(1^{h_1},2^{-h_1},3^{h_3},4^{-h_3}) \:,
\vspaceinarray\\ &&
M_i(1_{f_1}^{h_1},2_{f_2}^{-h_1},3_{f_3}^{h_3},4_{f_4}^{-h_3},5^{h_5}) =
\\ \nn && \qquad
2\,e^2\,g^3\,C_{f_1f_2}^{h_\ell, h_1}\,\delta_{f_3f_4}\,\frac{\i}{s}\,
A_i(1^{h_1},2^{-h_1},3^{h_3},4^{-h_3},5^{h_5}) \:,
\vspaceinarray\\ &&
\widetilde{M}(1_{f_1}^{h_1},2_{f_2}^{-h_1},3^{h_3},4^{-h_4}) =
\\ \nn && \qquad
2\,e^2\,g^2\,C_{f_1f_2}^{h_\ell, h_1}\,\frac{\i}{s}\,
A(1^{h_1},2^{-h_1},3^{h_3},4^{-h_3}) \:,
\vspaceinarray\\ &&
\widetilde{M}_i(1_{f_1}^{h_1},2_{f_2}^{-h_1},3^{h_3},4^{-h_3},5^{h_5}) =
\\ \nn && \qquad
2\,e^2\,g^3\,C_{f_1f_2}^{h_\ell, h_1}\,\frac{\i}{s}\,
A_i(1^{h_1},2^{-h_1},3^{h_3},4^{-h_3},5^{h_5}) \:,
\eeeq
with $s = Q^2 = (p_\ell+p_{\bar{\ell}})^2$.
The $C$ coefficients contain the electroweak couplings. If the vector
boson $V$ is $\gamma$ or $Z^0$ this coefficient is defined by
\beq
\label{CZ}
C_{f_1f_2}^{h_\ell, h_{f_1}}=
\left( - Q^{f_1} + v^{h_\ell}_\ell v^{h_{f_1}}_{f_1} \,{\cal P}_{Z}(s)
\right)\delta_{f_1f_2}\:, 
\eeq
where $f_1$, $f_2$ are the flavour indices of the quark antiquark pair 
that couples to the vector boson and
\beeq
&&
\label{ve}
v^-_\ell  = \frac{ -1 + 2\sin^2 \theta_W}{\sin 2 \theta_W } \,,
\qquad\;
v^+_\ell  = \frac{ 2 \sin^2 \theta_W }{  \sin 2 \theta_W } \,,
\vspaceinarray\\ &&
\label{vf}
v^-_f  = \frac{ \pm 1 - 2 Q_f\sin^2 \theta_W }{  \sin 2 \theta_W } \,,
\quad
v^+_f = -\frac{ 2 Q_f \sin^2 \theta_W }{ \sin 2 \theta_W }  \,
\eeeq
are the left- and right-handed couplings of leptons and quarks to neutral
gauge bosons.
In Eqs.~(\ref{ve},\ref{vf}) $\theta_W$ denotes the Weinberg angle, $Q_f$
is the electric charge of the quark of flavor $f$ in units of $e$ and
the two signs in Eq.~(\ref{vf}) correspond to up (+) and down ($-$) type
quarks. The coupling $C$ contains the ratio of the $Z^0$ and
photon propagators,
\beq
{\cal P}_{Z}(s) = {s \over s - M_Z^2 + \i \Gamma_Z \, M_Z}\;,
\eeq
where $M_Z$ and $\Gamma_Z$ are the mass and width of the $Z^0$.

If the vector boson $V$ is a $W^+$ or a $W^-$, then the couplings take
the form
\beq
\label{CW}
C_{f_1f_2}^{h_\ell,h_{f_1}}= v^{h_\ell}_\ell v^{h_{f_1}}_{f_1}
\,{\cal P}_{W}(s)\, \delta_{\tilde{f}_1f_2}\:,
\eeq
where $\tilde{f}_1$ denotes the partner of quark $f_1$ in the SU(2)$_L$
doublet and, for the sake of simplicity, we set the Kobayashi-Maskawa
mixing matrix to unity.  In Eq.~(\ref{CW}) the left- and right-handed
couplings differ from the corresponding expressions in
Eqs.~(\ref{ve},\ref{vf}):
\beq
v^-_\ell  = v^-_f = \frac{1}{2\sqrt2 \sin\theta_W } \,,
\qquad v_\ell^+ = v_f^+ = 0 \,.
\eeq
In this case ${\cal P}_{W}(s)$ denotes the ratio of the $W^{\pm}$ and
photon propagators,
\beq
{\cal P}_{W}(s) = {s \over s - M_W^2 + \i \Gamma_W \, M_W}\:,
\eeq
where $M_W$ and $\Gamma_W$ are the mass and width of the $W^{\pm}$.

\subsection{Four-parton color subamplitudes}
\label{a4parton}

In this subsection, we present all four-parton color subamplitudes for
the helicity configuration $h_q=+$ and $h_\ell=+$. The amplitudes for the
reversed helicity configurations can be obtained from these amplitudes
by applying parity operation $P$, which amounts to making the substitutions
$\a ij\equiv\la k_i^-|k_j^+ \ra\lra\b ji\equiv\la k_j^+|k_i^-\ra$.
The amplitudes when only the lepton helicities are reversed can be
obtained simply by exchanging the lepton labels and flipping the lepton
helicity in the coupling factors $C_{f_1f_2}^{h_\ell,h_{f_1}}$
We use the notation
\beeq
&&
\la i|lm\dots|j\ra\equiv k_l^\mu k_m^\nu 
\la k_i^-|\gamma_\mu\gamma_\nu\dots|k_j^\pm\ra\:,
\vspaceinarray
\\ &&
[i|lm\dots|j]\equiv k_l^\mu k_m^\nu 
\la k_i^+|\gamma_\mu\gamma_\nu\dots|k_j^\pm\ra\:,
\vspaceinarray
\\ &&
\la i|(l+m)\dots|j\ra \equiv (k_l^\mu+k_m^\mu)\dots
\bra{k_i^-}\,\gamma_\mu\dots\ket{k_j^\pm}\:,
\vspaceinarray
\\ &&
[i|(l+m)\dots|j] \equiv (k_l^\mu+k_m^\mu)\dots
\bra{k_i^+}\,\gamma_\mu\dots\ket{k_j^\pm}\:,
\eeeq
and the two- and three-particle invariants $\s ij\equiv (k_i+k_j)^2$ and
$\t ijl\equiv (k_i+k_j+k_l)^2$.
Labels 5 and 6 refer to the positron and electron respectively.

The two-quark two-gluon amplitudes are as follows:
\begin{eqnarray}
&&
A(1_q^+,2_g^+,3_g^+,4_\qb^-) = 
-\frac{\a45^2 \b56}{\a12 \a23 \a34}\:,
\nexteq
A(1_q^+,2_g^-,3_g^-,4_\qb^-) = 
-\frac{\b16^2 \a56}{\b12 \b23 \b34}\:,
\nexteq
A(1_q^+,2_g^+,3_g^-,4_\qb^-) = 
- \frac{\a31 \b12 \a45 \c3126}{\a12 \s23 \t123}
%\nextline
\vspaceinarray \\ \nn && \:
+ \frac{\a34 \b42 \b16 \c5342}{\b34 \s23 \t234}
+ \frac{\c5342 \c3126}{\a12 \b34 \s23}\:,
\nexteq
A(1_q^+,2_g^-,3_g^+,4_\qb^-) = 
\frac{\b13^2 \a45 \c2136}{\b12 \s23 \t123}
\nextline
- \frac{\a24^2 \b16 \c5243}{\a34 \s23 \t234}
- \frac{\b13 \a24 \b16 \a45}{\b12 \a34 \s23}\:.
\end{eqnarray}

The four-quark amplitudes are as follows:
\begin{eqnarray}
&&
A(1_q^+,2_\qb^-,3_Q^+,4_\Qb^-) = 
\nextline
- \Bigg(\frac{\b13 \a52 \c4136}{\t134 \s34}
+ \frac{\a42 \b61 \c5243}{\t234 \s34}\Bigg)\:,
\nexteq
A(1_q^+,2_\qb^-,3_Q^-,4_\Qb^+) = A(1_q^+,2_\qb^-,4_Q^+,3_\Qb^-) \:.
\end{eqnarray}

\subsection{Five-parton color subamplitudes}
\label{a5parton}

In this subsection, we present all five-parton color subamplitudes for
the helicity configuration $h_q=+$ and $h_\ell=+$. The amplitudes for the
remaining helicity configurations can be obtained from these amplitudes
as in the $n=4$ case.
Labels 6 and 7 refer to the positron and electron respectively.

First we list the two-quark three-gluon amplitudes:

%\beq
%A(1_q^+,2_g^+,3_g^+,4_g^+,5_\qb^-) =
%-\frac{\a65^2 \b67}{\a12 \a23 \a34 \a45}\:,
%\eeq

\onecolumn
%\footnotesize

\begin{eqnarray}
&&
A(1_q^+,2_g^+,3_g^+,4_g^+,5_\qb^-) =
-\frac{\a65^2 \b67}{\a12 \a23 \a34 \a45}\:,
\nexteq
A(1_q^+,2_g^+,3_g^+,4_g^-,5_\qb^-) =
\frac{\a65 \c4567}{\a23 \a34 \b34 \t567}
\left(\frac{\b23 \c4231}{\t234} + \frac{\c4123}{\a12}\right) 
\nextline
+ \frac{\c4567 \c6543}{\a12 \a23 \a34 \b34 \b45}
- \frac{\c4127 \c6543 \a45 \b53}{\a12 \a24 \b45 \s34 \t345} 
- \frac{\b17 \c6172 \a45^2 \b53^2}{\b45 \a42 \s34 \t345 \t167} 
\nextline
+ \frac{\b17 \a45 \b53}{\a23 \b34 \b45 \t167}
\left(\frac{\c6172}{\a34} + \frac{\c6173}{\a24}\right)
- \frac{\b17 \a45 \b23}{\a23 \b34 \t234 \t167}
\left(\frac{\c6172 \a24}{\a34} + \c6173\right)\:,
\nexteq
A(1_q^+,2_g^+,3_g^-,4_g^+,5_\qb^-) =
  \frac{\a31 \b12 \c3124 \c3567 \a65}{\a12 \a34 \s23 \t123 \t567} 
+ \frac{\a31 \b12 \c3127 \a65 \a35}{\a12 \a34 \a45 \s23 \t123}
\nextline
- \frac{\c3127 \c6172 \a35^2}{\a12 \a23 \a34 \a45 \b23 \t345} 
+ \frac{\c3127 \c6534 \a35 \b42}{\b23 \a12 \a23 \s34 \t345}
- \frac{\b42^2 \b17 \c61{7)(2+4}3 \a35}{\s23 \s34 \t234 \t167} 
\nextline
+ \frac{\b42 \c3567 \a65}{\s23 \s34 \t567}
 \left(\frac{\b42 \c3241 }{\t234} - \frac{\c3124}{\a12}\right)
+ \frac{\b17 \c6172 \a35^2}{\s23 \t345 \t167}
 \left(\frac{\c3542}{\a34 \a45} + \frac{\b42 \b54}{\s34}\right)\:,
%\nexteq
%A(1_q^+,2_g^+,3_g^-,4_g^+,5_\qb^-) =
%\frac{\b42^2 \c3241 \c3567 \a65}{\s23 \s34 \t234 \t567} 
%+ \frac{\b42 \c3124 \c3567 \a65}{\b23 \a12 \a23 \s34 \t567}
%\nextline
%+ \frac{\a31 \b12 \c3124 \c3567 \a65}{\a12 \a34 \s23 \t123 \t567} 
%+ \frac{\a31 \b12 \c3127 \a65 \a35}{\a12 \a34 \a45 \s23 \t123}
%\nextline
%- \frac{\c3127 \c6172 \a35^2}{\a12 \a23 \a34 \a45 \b23 \t345} 
%+ \frac{\c3127 \c6534 \a35 \b42}{\b23 \a12 \a23 \s34 \t345}
%\nextline
%- \frac{\b42^2 \b17 \c61{7)(2+4}3 \a35}{\s23 \s34 \t234 \t167} 
%+ \frac{\b17 \c6172 \a35^2}{\s23 \t345 \t167}
% \left(\frac{\c3542}{\a34 \a45} + \frac{\b42 \b54}{\s34}\right)\:,
\nexteq
A(1_q^+,2_g^-,3_g^+,4_g^+,5_\qb^-) =
\frac{\b43^2 \c2341 \c2567 \a65}{\s23 \s34 \t234 \t567}
+ \frac{\b13 \c2341 \c2567 \a65}{\b12 \a34 \a42 \s23 \t567}
\nextline
- \frac{\b13^2 \c2134 \c2567 \a65}{\b12 \a24 \s23 \t123 \t567}
- \frac{\b13^2 \c2137 \a65 \a25}{\b12 \a24 \a45 \s23 \t123}
- \frac{\b17 \a25 \b43^2 \c61{7)(3+4}2}{\s23 \s34 \t234 \t167} 
\nextline
+ \frac{\b13 \b17 \a25}{\b12 \s23 \a24 \t345}
\left(\c6543 \left(\frac{\a25}{\a45} - \frac{\a32}{\a34}\right)
  - \c6534 \frac{\a42}{\a34}\right) 
\nextline
+ \frac{\b17 \c6173 \a25}{\a24 \s23 \t345 \t167}
\left(\c2543 \left(\frac{\a25}{\a45}-\frac{\a32}{\a34}\right)
  -\c2534 \frac{\a42}{\a34}\right)\:,
\nexteq
A(1_q^+,2_g^+,3_g^-,4_g^-,5_\qb^-) =
\frac{\b12 [2|(3+4)(5+6)|7] \a65}{\s23 \t567}
 \left(\frac{\a43^2}{\s34 \t234} - \frac{\a31}{\b34 \b42 \a12}\right)
- \frac{\a31^2 \b12^2 \c4567 \a65}{\a12 \b24 \s23 \t123 \t567}
\nextline
+ \frac{\a31 \b12 \c3127 \c6542}{\a12 \b24 \b45 \s23 \t123} 
+ \frac{\c3127 \c6172}{\a12 \b34 \b45 \s23}
+ \frac{\b17 \c6172 \c3542}{\b34 \b45 \s23 \t167}
\nextline
- \frac{\b17 \c6172 \c5342 \a43^2}{\s23 \s34 \t234 \t167}\:,
\nexteq
A(1_q^+,2_g^-,3_g^-,4_g^+,5_\qb^-) =
\frac{\a23^2 \b14 [4|(2+3)(5+6)|7]\a65}{\s23 \s34 \t234 \t567}
\nextline
- \frac{\b14 \c3567 \a65}{\b42 \s34 \t123 \t567}
\left(\frac{\b42 \c2134 + \b43 \c3124}{\b23} 
  - \frac{\b14 \c3124}{\b12}\right) 
\nextline
- \frac{\b14 \a65 \a35}{\b42 \a45 \s34 \t123}
\left(\frac{\b42 \c2137 + \b43 \c3127}{\b23}
  - \frac{\b14 \c3127}{\b12}\right) 
+ \frac{\b14 \b17 \c6534 \a35^2}{\b12 \b24 \a45 \s34 \t345}
\nextline
- \frac{\b17 \c6174 \c2534 \a35^2}{\b42 \a45 \s34 \t345 \t167}
- \frac{\b17 \c6174 \c5234 \a35}{\a45 \b42 \b23 \s34 \t167}
- \frac{\b17 \c6174 \c5234 \a23^2}{\s23 \s34 \t234 \t167}\:,
\nexteq
A(1_q^+,2_g^-,3_g^+,4_g^-,5_\qb^-) =
P\,A(5_q^+,4_g^+,3_g^-,2_g^+,1_\qb^-)\st{6\lra7}\,,\quad
%\frac{\a42^2 \b13 [3|(2+4)(5+6)|7] \a65}{\s23 \s34 \t234 \t567}
%- \frac{\a42 \b13^2 \c4567 \a65}{\b12 \s23 \s34 \t567}
%\nextline
%+ \frac{\b13^3 \a21 \c4567 \a65}{\b12 \b34 \s23 \t123 \t567}
%- \frac{\b13^2 \c2137 \c6543}{\b12 \b34 \b45 \s23 \t123}
%- \frac{\b17 \c6173 \a42^2 \c5243}{\s23 \s34 \t234 \t167} 
%\nextline
%- \frac{\b13 \b17 \c6543}{\b12 \s23 \t345}
%\left (\frac{\a45 \a42}{\s34} - \frac{\c2543}{\b34 \b45}\right)
%- \frac{\b17 \c6173 \c2543}{\s23 \t345 \t167}
%\left(\frac{\a45 \a42}{\s34} - \frac{\c2543}{\b34 \b45}\right)\:,
%\nexteq
A(1_q^+,2_g^-,3_g^-,4_g^-,5_\qb^-) =
P\,A(5_q^+,4_g^+,3_g^+,2_g^+,1_\qb^-)\st{6\lra7}\,.
%\frac{\b17^2 \a67}{\b12 \b23 \b34 \b45}\:,
\end{eqnarray}

The four-quark one-gluon amplitudes have the form:
\begin{eqnarray}
&&
A_1(1_q^+,2_\qb^-,3_Q^+,4_\Qb^-,5_g^+) = 
- \frac{\b15 \c4153 \c4267 \a62}{\a45 \s15 \s34 \t267}
- \frac{\b17 \c6175 \a42^2 \b23}{\a45 \s34 \t234 \t167}
\nextline
- \frac{\c4157 \c6243 \a42}{\a15 \a54 \s34 \t234}
+ \frac{\b53 \c4351 \c4267 \a62}{\a45 \s34 \t345 \t267}
+ \frac{\b17 \c61{7)(3+5}4 \b35 \a42}{\a45 \s34 \t345 \t167}\:,
\nexteq
A_1(1_q^+,2_\qb^-,3_Q^+,4_\Qb^-,5_g^-) = 
\frac{\b13^2 \a51 \c4267 \a62}{\b35 \s15 \s34 \t267}
+ \frac{\b17 \c6173 \c5243 \a42}{\b35 \s34 \t234 \t167}
\nextline
- \frac{\b13 \b17 \c6243 \a42}{\b15 \b53 \s34 \t234}
+ \frac{\b13 \a54 [3|(4+5)(2+6)|7] \a62}{\b35 \s34 \t345 \t267}
- \frac{\b17 \c6173 \a54 \c2453}{\b35 \s34 \t345 \t167}\:,
\nexteq
A_1(1_q^+,2_\qb^-,3_Q^-,4_\Qb^+,5_g^+) = 
A_1(1_q^+,2_\qb^-,4_Q^+,3_\Qb^-,5_g^+)\:,
%- \frac{\b15 \c3154 \c3267 \a62}{\a35 \s15 \s34 \t267}
%- \frac{\b17 \c6175 \a32^2 \b24}{\a35 \s34 \t243 \t167}
%\nextline
%- \frac{\c3157 \c6234 \a32}{\a15 \a53 \s34 \t243}
%+ \frac{\b54 \c3451 \c3267 \a62}{\a35 \s34 \t345 \t267}
%+ \frac{\b17 \c61{7)(4+5}3 \b45 \a32}{\a35 \s34 \t345 \t167}\:,
%\nexteq
\qquad
A_1(1_q^+,2_\qb^-,3_Q^-,4_\Qb^+,5_g^-) = 
A_1(1_q^+,2_\qb^-,4_Q^+,3_\Qb^-,5_g^-)\:,
%\frac{\b14^2 \a51 \c3267 \a62}{\b45 \s15 \s34 \t267}
%+ \frac{\b17 \c6174 \c5234 \a32}{\b45 \s34 \t243 \t167}
%\nextline
%- \frac{\b14 \b17 \c6234 \a32}{\b15 \b54 \s34 \t243}
%+ \frac{\b14 \a53 [4|(3+5)(2+6)|7] \a62}{\b45 \s34 \t435 \t267}
%- \frac{\b17 \c6174 \a53 \c2354}{\b45 \s34 \t435 \t167}\:,
\nexteq
A_2(1_q^+,2_\qb^-,3_Q^+,4_\Qb^-,5_g^+) = 
- \frac{\b53 \c4351 \c4267 \a62}{\a45 \s34 \t345 \t267}
- \frac{\b17 \c6173 \b25 \a42^2}{\a45 \s25 \s34 \t167} 
\nextline
- \frac{\b13 \c4135 \c4267 \a62}{\a45 \s34 \t134 \t267}
- \frac{\b13 \c4137 \a42 \a62}{\a45 \a52 \s34 \t134}
- \frac{\b17 \c61{7)(3+5}4 \b35 \a42}{\a45 \s34 \t345 \t167}\:,
\nexteq
A_2(1_q^+,2_\qb^-,3_Q^+,4_\Qb^-,5_g^-) = 
- \frac{\b13 \a54 [3|(4+5)(2+6)|7] \a62}{\b35 \s34 \t345 \t267}
+ \frac{\b13^2 \a41 \c5267 \a62}{\b35 \s34 \t134 \t267} 
\nextline
+ \frac{\b17 \c6173 \a54 \c2453}{\b35 \s34 \t345 \t167}
- \frac{\b13 \c4137 \c6253}{\b35 \b52 \s34 \t134} 
+ \frac{\b17 \c6173 \a52 \c4253}{\b35 \s25 \s34 \t167}\:,
\nexteq
A_2(1_q^+,2_\qb^-,3_Q^-,4_\Qb^+,5_g^+) = 
A_2(1_q^+,2_\qb^-,4_Q^+,3_\Qb^-,5_g^+)\:,
%- \frac{\b54 \c3451 \c3267 \a62}{\a35 \s34 \t345 \t267}
%- \frac{\b14 \c3147 \a62 \a32}{\a35 \a52 \s34 \t143}
%\nextline
%- \frac{\b17 \c61{7)(4+5}3 \b45 \a32}{\a35 \s34 \t345 \t167}
%- \frac{\b17 \c6174 \b25 \a32^2}{\a35 \s25 \s34 \t167}
%- \frac{\b14 \c3145 \c3267 \a62}{\a35 \s34 \t143 \t267}\:,
%\nexteq
\qquad
A_2(1_q^+,2_\qb^-,3_Q^-,4_\Qb^+,5_g^-) = 
A_2(1_q^+,2_\qb^-,4_Q^+,3_\Qb^-,5_g^-)\:,
%- \frac{\b14 \a53 [4|(3+5)(2+6)|7] \a62}{\b45 \s34 \t435 \t267}
%+ \frac{\b14^2 \a31 \c5267 \a62}{\b45 \s34 \t143 \t267}
%\nextline
%+ \frac{\b17 \c6174 \a53 \c2354}{\b45 \s34 \t435 \t167}
%- \frac{\b14 \c3147 \c6254}{\b45 \b52 \s34 \t143}
%+ \frac{\b17 \c6174 \a52 \c3254}{\b45 \s25 \s34 \t167}\:,
\nexteq
A_3(1_q^+,2_\qb^-,3_Q^+,4_\Qb^-,5_g^+) = 
- \frac{\b53 \c4351 \c4267 \a62}{\a45 \s35 \t345 \t267}
- \frac{\b17 \c61{7)(3+5}4 \b35 \a42}{\a45 \s35 \t345 \t167}\:,
\nexteq
A_3(1_q^+,2_\qb^-,3_Q^-,4_\Qb^+,5_g^-) = 
- \frac{\b14 \a53 [4|(3+5)(2+6)|7] \a62}{\b45 \s35 \t435 \t267}
+ \frac{\b17 \c6174 \a53 \c2354}{\b45 \s35 \t435 \t167}\:,
\nexteq
A_4(1_q^+,2_\qb^-,3_Q^+,4_\Qb^-,5_g^-) = 
\frac{\b13 \a54 [3|(4+5)(2+6)|7] \a62}{\b35 \s45 \t345 \t267}
- \frac{\b17 \c6173 \a54 \c2453}{\b35 \s45 \t345 \t167}\:,
\nexteq
A_4(1_q^+,2_\qb^-,3_Q^-,4_\Qb^+,5_g^+) = 
\frac{\b54 \c3451 \c3267 \a62}{\a35 \s45 \t345 \t267}
+ \frac{\b17 \c61{7)(4+5}3 \b45 \a32}{\a35 \s45 \t345 \t167}
\end{eqnarray}

\twocolumn
%\normalsize

\section{Matrix elements}

In this appendix we present analytic formulas for the
color-correlated four-parton Born-level matrix elements and for the
four-, five-parton Born-level matrix elements. The calculation of the
color-correlated four-parton amplitudes is a straightforward application
of color algebra and the four-parton helicity amplitudes. However, to our
knowledge these results were not published previously. The uncorrelated
color sum was first calculated in Ref.~\cite{a5parton}. We present
our results in terms of the color subamplitudes of Appendix A.  It is a
new feature of the matrix elements in this appendix that they are given
in terms of group independent functions and eigenvalues of the quadratic
Casimir operators of the underlying gauge group.

Having the helicity amplitudes at our disposal, we calculate the squared
matrix elements summed over final state colors without and with
color-correlation:
\beeq
&&
\left|\M_n (1,\ldots,n)\right|^2 =
\\ \nn && \qquad
\nbra{1,\dots,n} \nket{1,\dots,n}\:,\quad
n=4,5\:,
\vspaceinarray \\ &&
\left|\M^{i,j}_4 (1,\ldots,4)\right|^2 =
\\ \nn && \qquad
\mbra{1^{h_1},\dots,4^{h_4}}
%\,{\bom T}_i \cdot {\bom T}_j\,{\bom H}_{i,j}^{hh'}\,
\,{\bom T}_i \cdot {\bom T}_j \,
\mket{1^{h'_1},\dots,4^{h'_4}}\:,
\eeeq
where in the latter case we leave the helicity indices explicit so that
both correlated and uncorrelated helicity summation is possible.
(Although we did not show the flavor indices, the flavor summation is
also left out, as will become clear later.)
In the correlated case we have to insert the helicity matrix
(see Eq.~(\ref{dipole}))
\beq
{\bom H}_{i,j}^{hh'} =
\delta_{h_1h'_1}\dots \bra{h_i}\,{\bom V}_{i,j}\,\ket{h'_i}
\dots \delta_{h_nh'_n}\:,
\eeq
and in the uncorrelated case
\beq
{\bom H}_{i,j}^{hh'} =
\delta_{h_1h'_1}\dots \delta_{h_ih'_i} \dots \delta_{h_nh'_n}\:.
\eeq

We evaluate the color sum in such a way that the
matrix elements are given as polynomial expressions of the Casimir
invariants of the gauge group with group independent kinematical
coefficients. In addition to the usual quadratic Casimirs $C_F$ and
$C_A$, we shall also use a cubic Casimir $C_3$ that is defined as
\beq
C_3=\sum_{a,b,c=1}^{N_A} {\rm Tr}(t^a t^b t^c) {\rm Tr}(t^c t^b t^a)\:.
\eeq
In the following subsections we list the explicit formulas for
$|\M_4|^2$, $|\M_4^{ij}|^2$ and $|\M_5|^2$.

\subsection{Four-parton color-summed matrix elements}

In this subsection, we give explicit formulas for the color-summed Born
matrix elements for four final state partons.
There are four different cases: the two-quark two-gluon process and three
four-fermion processes (two unequal flavor quark pairs, two equal flavor
quark pairs and the two-quark two-gluino production).
The color summation is straightforward in each cases, we simply list the
results:
\beeq
&&
\left|{\cal M}_4(1_q, 2_g, 3_g, 4_\qb)\right|^2 = \Nc C_F^2
\\ \nn &&\quad
\times\Big\{
|m(1_{f_1},2,3,4_{f_4})|^2 + |m(1_{f_1},3,2,4_{f_4})|^2
\\ \nn &&\qquad
+2\,{\rm Re}\left(m(1_{f_1},2,3,4_{f_4})m(1_{f_1},3,2,4_{f_4})^*\right)
\\ \nn &&\qquad
-x\,{\rm Re}\big(m(1_{f_1},2,3,4_{f_4})m(1_{f_1},3,2,4_{f_4})^*\big) \Big\}\:,
\nexteq
\left |{\cal M}_4(1_q, 2_\qb, 3_Q, 4_\Qb)\right|^2= \Nc C_F^2
\\ \nn &&\quad
\times\Big\{
-2\,{\rm Re}\big({\overline M}(1_{f_1},2_{f_2},3_{f_3},4_{f_4})
                 {\overline M}(1_{f_1},4_{f_4},3_{f_3},2_{f_2})^*\big)
\\ \nn &&\qquad
+x \,{\rm Re}\big({\overline M}(1_{f_1},2_{f_2},3_{f_3},4_{f_4})
                  {\overline M}(1_{f_1},4_{f_4},3_{f_3},2_{f_2})^*\big)
\\ \nn &&\qquad
+y\,\left(\left|{\overline M}(1_{f_1},2_{f_2},3_{f_3},4_{f_4})\right|^2\right.
\\ \nn &&\qquad\quad\,
   \left.+\left|{\overline M}(1_{f_1},4_{f_4},3_{f_3},2_{f_2})\right|^2\right)
\Big\}\:,
\nexteq
\left |{\cal M}_4(1_q, 2_\qb, 3_\gt, 4_\gt)\right|^2 =
\\ \nn &&\qquad
\Nc C_F^2\,x \left|\widetilde{M}(1_{f_1},2_{f_2},3,4)\right|^2\:,
\eeeq
where $x$ and $y$ are ratios of the quadratic Casimirs (see
Eq.~\ref{xydef}) and ${\overline M}(1_{f_1},2_{f_2},3_{f_3},4_{f_4})$
is defined by
\beeq
&&
{\overline M}(1_{f_1},2_{f_2},3_{f_3},4_{f_4})=
\\ \nn &&\qquad
M(1_{f_1},2_{f_2},3_{f_3},4_{f_4})+M(3_{f_3},4_{f_4},1_{f_1},2_{f_2})\:.
\eeeq

\subsection{Four-parton color-correlated, color-summed matrix elements}

In this subsection, we give explicit formulas for the color-correlated,
color-summed Born matrix elements for four final state partons.
We consider those four cases as in the previous subsection.
The color summation is again fairly straightforward, therefore, we only
record the results.

For the $V\to q\qb gg$ subprocess
\beeq
&&
\left|{\cal M}_4^{ik}(1_q, 2_g, 3_g, 4_\qb)\right|^2 =
\\ \nn &&\qquad
-\frac{\Nc C_F^3}2 \left\{ M_0^{ik} + x M_x^{ik}
+ x^2 M_{xx}^{ik}\right\}\,,
\eeeq
where the non-vanishing elements of the matrices $M_0^{ik}$, $M_x^{ik}$,
$M_{xx}^{ik}$ are given by
\beeq
&&
M_0^{12} = 2(S_1+S_2+S_3)\:,
\vspaceinarray \\ &&
M_x^{12} = -2S_1-2S_2-3 S_3\:,
\\ \nn &&
M_x^{13} = M_x^{14} = M_x^{23} = M_x^{24} = S_1+S_2+S_3\:,
\vspaceinarray \\ &&
M_{xx}^{12} = \frac12 (S_1+S_2+2 S_3)\:,
\\ \nn &&
M_{xx}^{13} = M_{xx}^{24} = -\frac12 (S_2+S_3)\:,
\\ \nn &&
M_{xx}^{14} = M_{xx}^{23} = -\frac12 (S_1+S_3)\:,
\\ \nn &&
M_{xx}^{34} =  \frac12 (S_1+S_2)
\eeeq
and the $S_i$ functions are defined by
\beeq
&&
S_1 = 
m(1_{f_1}^{h_1},2^{h_2},3^{h_3},4_{f_4}^{h_4})^*
m(1_{f_1}^{h'_1},2^{h'_2},3^{h'_3},4_{f_4}^{h'_4})\:,
\vspaceinarray \\ &&
S_2 = 
m(1_{f_1}^{h_1},3^{h_3},2^{h_2},4_{f_4}^{h_4})^*
m(1_{f_1}^{h'_1},3^{h'_3},2^{h'_2},4_{f_4}^{h'_4})\:,
\vspaceinarray \\ &&
S_3 = 
m(1_{f_1}^{h_1},2^{h_2},3^{h_3},4_{f_4}^{h_4})^*
m(1_{f_1}^{h'_1},3^{h'_3},2^{h'_2},4_{f_4}^{h'_4})
\\ \nn && \quad\;\,
+m(1_{f_1}^{h_1},3^{h_3},2^{h_2},4_{f_4}^{h_4})^*
m(1_{f_1}^{h'_1},2^{h'_2},3^{h'_3},4_{f_4}^{h'_4})\:.
\eeeq

In the case of four-quark production
\beeq
&&
\left|{\cal M}_4^{ik}(1_q, 2_\qb, 3_Q, 4_\Qb)\right|^2 =
\\ \nn &&\qquad
-\frac{\Nc C_F^3}2 
\Big\{ M_0^{ik} + x M_x^{ik} + y M_y^{ik} + z M_z^{ik}
\\ \nn &&\qquad\qquad\qquad
  + x^2 M_{xx}^{ik} + xy M_{xy}^{ik}\Big\}\,,
\eeeq
where the non-zero element of the matrices $M_0^{ik}$, $M_x^{ik}$,
$M_y^{ik}$, $M_z^{ik}$, $M_{xx}^{ik}$, $M_{xy}^{ik}$ are the following
ones:
\beeq
&&
M_0^{12} = M_0^{14} = M_0^{23} = M_0^{34} = -2S_3\:,
\\ \nn &&
M_0^{13} = M_0^{24} = 2S_3\:,
\vspaceinarray \\ &&
M_x^{12} = M_x^{14} = M_x^{23} =  M_x^{34} = 2S_3\:,
\\ \nn &&
M_x^{13} = M_x^{24} = -3S_3\:,
\vspaceinarray \\ &&
M_y^{12} =  M_y^{34} = 2S_1\:,
\qquad%\\ \nn &&
M_y^{14} = M_y^{23} = 2S_2\:,
\vspaceinarray \\ &&
M_z^{12} =  M_z^{34} = 2S_2\:,
\qquad%\\ \nn &&
M_z^{14} = M_z^{23} = 2S_1\:,
\\ \nn &&
M_z^{13} = M_z^{24} = -2(S_1+S_2)\:,
\vspaceinarray \\ &&
M_{xx}^{12} = M_{xx}^{14} = M_{xx}^{23} = M_{xx}^{34} = -\frac12S_3\:,
\\ \nn &&
M_{xx}^{13} = M_{xx}^{24} = S_3\:,
\vspaceinarray \\ &&
M_{xy}^{12} =  M_{xy}^{34} = -S_1\:,
\qquad%\\ \nn &&
M_{xy}^{14} = M_{xy}^{23} = -S_2\:,
\\ \nn &&
M_{xy}^{13} = M_{xy}^{24} =  S_1+S_2\:.
\eeeq
For this case the $S_i$ functions are defined as follows:
\beeq
&&
S_1 = 
{\overline M}(1_{f_1}^{h_1},2_{f_2}^{h_2},3_{f_3}^{h_3},4_{f_4}^{h_4})^*
{\overline M}(1_{f_1}^{h'_1},2_{f_2}^{h'_2},3_{f_3}^{h'_3},4_{f_4}^{h'_4})\:,
\vspaceinarray \\ &&
S_2 = 
{\overline M}(1_{f_1}^{h_1},4_{f_4}^{h_4},3_{f_3}^{h_3},2_{f_2}^{h_2})^*
{\overline M}(1_{f_1}^{h'_1},4_{f_4}^{h'_4},3_{f_3}^{h'_3},2_{f_2}^{h'_2})\:,
\vspaceinarray \\ &&
S_3 = 
{\overline M}(1_{f_1}^{h_1},2_{f_2}^{h_2},3_{f_3}^{h_3},4_{f_4}^{h_4})^*
{\overline M}(1_{f_1}^{h'_1},4_{f_4}^{h'_4},3_{f_3}^{h'_3},2_{f_2}^{h'_2})
\\ \nn && \quad\;\,
+{\overline M}(1_{f_1}^{h_1},4_{f_4}^{h_4},3_{f_3}^{h_3},2_{f_2}^{h_2})^*
{\overline M}(1_{f_1}^{h'_1},2_{f_2}^{h'_2},3_{f_3}^{h'_3},4_{f_4}^{h'_4})\:,
\eeeq
where 
\beeq
&&
{\overline M}(1_{f_1},2_{f_2},3_{f_3},4_{f_4}) =
\\ \nn && \qquad
M(1_{f_1},2_{f_2},3_{f_3},4_{f_4})+M(1_{f_1},4_{f_4},3_{f_3},2_{f_2})\:.
\eeeq
Finally, for the $V\to q\qb\gt\gt$ subprocess
\beeq
&&
\left|{\cal M}_4^{ik}(1_q, 2_\qb, 3_\gt, 4_\gt)\right|^2 =
\\ \nn &&\qquad
-\frac{\Nc C_F^3}2
\left\{x \widetilde{M}_x^{ik} + x^2 \widetilde{M}_{xx}^{ik}\right\}\,,
\eeeq
where the non-vanishing elements of the matrices $\widetilde{M}_x^{ik}$
and $\widetilde{M}_{xx}^{ik}$ are given by
\beeq
&&
\widetilde{M}_x^{12} = 
2\left|\widetilde{M}(1_{f_1},2_{f_2},3_{f_3},4_{f_4})\right|^2\:,
\vspaceinarray \\ &&
\widetilde{M}_{xx}^{12} =  -\widetilde{M}_{xx}^{34} =
-\left|\widetilde{M}(1_{f_1},2_{f_2},3_{f_3},4_{f_4})\right|^2 \:,
%\\ \nn &&
%\widetilde{M}_{xx}^{34} =
%\left|\widetilde{M}(1_{f_1},2_{f_2},3_{f_3},4_{f_4})\right|^2
\\ \nn &&
\widetilde{M}_{xx}^{13} = \widetilde{M}_{xx}^{14} = 
\widetilde{M}_{xx}^{23} = \widetilde{M}_{xx}^{24} =
\frac12 \left|\widetilde{M}(1_{f_1},2_{f_2},3_{f_3},4_{f_4})\right|^2 \:.
\eeeq

\subsection{Five-parton color-summed matrix elements}

In this subsection, we give explicit formulas for the color-summed Born
matrix elements for five final state partons.  There are again four
different cases: the two-quark three-gluon process, the production of
two equal, or unequal flavor quark pairs plus a gluon and the two-quark
two-gluino one-gluon production.

In the case of the two-quark three-gluon process the color summation is
straightforward and leads to the following expression:
\beeq
&&
\left|{\cal M}_5(1_q, 2_g, 3_g, 4_g, 5_\qb)\right|^2 =
\\ \nn && \quad\!
\Nc C_F^3 \left\{ M_0 - \frac{x}2 (M_1+2 M_0)
+ \frac{x^2}4 (M_0+M_1+M_2)\right\}\:,
\eeeq
where
\beeq
&&
M_0 = \bigg|\sum_{\{2,3,4\}} m(1_{f_1},2,3,4,5_{f_5})\bigg|^2\:,
\vspaceinarray \\ &&
M_2 = \sum_{\{2,3,4\}}\left| m(1_{f_1},2,3,4,5_{f_5})\right|^2\:,
\eeeq
and
\twocolumn[\hsize\textwidth\columnwidth\hsize\csname
@twocolumnfalse\endcsname
\beeq
&&
M_1 = -2\,M_2
\\ \nn && \qquad\;
-2\,{\rm Re}\!\sum_{\{2,3,4\}'}\!
\Bigg\{m(1_{f_1},2,3,4,5_{f_5})^*
\bigg(m(1_{f_1},2,4,3,5_{f_5})
     +m(1_{f_1},3,2,4,5_{f_5})
     -m(1_{f_1},4,3,2,5_{f_5})\bigg)\Bigg\}\:,
\eeeq
]
\noindent
with $\{2,3,4\}'$ denoting the cyclic permutations of the three labels
2, 3 and 4.

In the case of the four-quark one-gluon subprocesses we have to evaluate
the following color sums:
\beeq
&&
T_1^\dag T_1 = T_2^\dag T_2 = \Nc C_F^2 T_R \:,
\vspaceinarray \\ &&
T_1^\dag T_1(2\lra 4) = T_2^\dag T_2(1\lra 3) =
\\ \nn && \qquad
\Nc C_F^2 \left(C_F-\frac{1}{2} C_A\right) \:,
\vspaceinarray \\ &&
T_1^\dag T_1(1\lra 3) = T_2^\dag T_2(2\lra 4) =
\\ \nn && \qquad
\Nc C_F \left(C_F-\frac{1}{2} C_A\right)\left(C_F- C_A\right) \:,
\vspaceinarray \\ &&
T_1^\dag T_1(1\lra 3, 2\lra 4) = T_2^\dag T_2(1\lra 3, 2\lra 4) =
\\ \nn && \qquad
C_3-\Nc C_F T_R \frac{C_A}{2}\:,
\vspaceinarray \\ &&
T_1^\dag T_2 = \Nc C_F T_R  \left(C_F-\frac{1}{2} C_A\right)\:,
\vspaceinarray \\ &&
T_1^\dag T_2(1\lra 3) = T_1^\dag T_2(2\lra 4) =
\\ \nn && \qquad
\Nc C_F  \left(C_F-\frac{1}{2} C_A\right)^2\:,
\vspaceinarray \\ &&
T_1^\dag T_2(1\lra 3, 2\lra 4) = C_3\:.
\eeeq
Using these results, the square of the matrix element for any flavor
configuration can be written in the form:
\beeq
&&
\left|{\cal M}_5(1_q, 2_\qb, 3_Q, 4_\Qb, 5_g)\right|^2 =
\\ \nn && \;
\Nc C_F^3
\left\{ M_0 + x M_x + y M_y + z M_z + x^2 M_{xx} + xy M_{xy}\right\}
\:,
\eeeq
where we have introduced the ratio
\beq
z = \frac{C_3}{\Nc C_F^3}
\eeq
and the following abbreviations:
\beeq
&&
M_0 = B+C+E\:,
\vspaceinarray \\ &&
M_x = -\frac12(3C+2E+B)\:,
\vspaceinarray \\ &&
M_y = A+D\:,
\vspaceinarray \\ &&
M_z = F+G\:,
\vspaceinarray \\ &&
M_{xx} = \frac14(2C+E)\:,
\vspaceinarray \\ &&
M_{xy} = -\frac12(F+D)\:,
\eeeq
with the functions $A$, $B$, $C$, $D$, $E$, $F$ defined as
\beeq
&&
A = \sum_{\{1,3\}}\sum_{\{2,4\}}\sum_{i=1}^2|{\overline M}_i|^2\:,
\vspaceinarray \\ &&
B = -2{\rm Re}
\big({\overline M}_1{\overline M}_1(2\lra4)^*+{\overline M}_2{\overline
M}_2(1\lra3)^* 
\\ \nn && \qquad\qquad
+ (1\lra3,2\lra4)\big)\:,
\vspaceinarray \\ &&
C = -2{\rm Re}
\big({\overline M}_1{\overline M}_1(1\lra3)^*+{\overline M}_2{\overline
M}_2(2\lra4)^* 
\\ \nn && \qquad\qquad
+ (1\lra3,2\lra4)\big)\:,
\vspaceinarray \\ &&
D = 2{\rm Re}
\Big(\sum_{\{1,3\}}\sum_{\{2,4\}}{\overline M}_1{\overline M}_2^*\Big)\:,
\vspaceinarray \\ &&
E = -2 {\rm Re}\big(({\overline M}_1+{\overline M}_1(1\lra3,2\lra4))
\\ \nn && \qquad\qquad
\times({\overline M}_2(1\lra3)+{\overline M}_2(2\lra4))^*
%\\ &&
%E = -2 {\rm Re}\big(({\overline M}_1(1,2,3,4,5)+{\overline M}_1(3,4,1,2,5))
%({\overline M}_2(3,2,1,4,5)+{\overline M}_2(1,2,3,2,5))^*
\\ \nn && \qquad\qquad
+ ({\overline M}_1\lra {\overline M}_2)\big)\:,
\vspaceinarray \\ &&
F = 2 {\rm Re}\big({\overline M}_1{\overline M}_1(1\lra3,2\lra4)^*
\\ \nn && \qquad\qquad
+ {\overline M}_1(1\lra3){\overline M}_1(2\lra4)^*
\\ \nn && \qquad\qquad
+ ({\overline M}_1\lra {\overline M}_2)\big)\:,
\vspaceinarray \\ &&
G = 2 {\rm Re}\big({\overline M}_1{\overline M}_2(1\lra3,2\lra4)^*
\\ \nn && \qquad\qquad
+ {\overline M}_1(1\lra3){\overline M}_2(2\lra4)^*
\\ \nn && \qquad\qquad
+ ({\overline M}_1\lra {\overline M}_2)\big)\:,
\eeeq
where
\beeq
&&
{\overline M}_1(1_{f_1},2_{f_2},3_{f_3},4_{f_4},5_g)=
\\ \nn &&\qquad
M_1(1_{f_1},2_{f_2},3_{f_3},4_{f_4},5_g)
+M_3(3_{f_3},4_{f_4},1_{f_1},2_{f_2},5_g)\:,
\\ \nn &&
{\overline M}_2(1_{f_1},2_{f_2},3_{f_3},4_{f_4},5_g)=
\\ \nn &&\qquad
M_2(1_{f_1},2_{f_2},3_{f_3},4_{f_4},5_g)
+M_4(3_{f_3},4_{f_4},1_{f_1},2_{f_2},5_g)\:.
\eeeq

For the $V \to q\qb\gt\gt g$ process we have to calculate the following
products of the $\widetilde{T}_i$ colour factors:
\beeq
&&
\widetilde{T}_1^\dag \widetilde{T}_1 =
\widetilde{T}_2^\dag \widetilde{T}_2 = \Nc C_F^2 C_A\:,
\vspaceinarray \\ &&
\widetilde{T}_3^\dag \widetilde{T}_3 =
\widetilde{T}_4^\dag \widetilde{T}_4 = \Nc C_F C_A^2\:,
\vspaceinarray \\ &&
\widetilde{T}_1^\dag \widetilde{T}_2 =                  
\Nc C_F C_A \left(C_F-\frac{1}{2} C_A\right)\:,
\vspaceinarray \\ &&
\widetilde{T}_1^\dag \widetilde{T}_3 =
\widetilde{T}_2^\dag \widetilde{T}_4 =
-\Nc C_F \frac{C_A^2}{4}\:,
\vspaceinarray \\ &&
\widetilde{T}_1^\dag \widetilde{T}_4 =
\widetilde{T}_2^\dag \widetilde{T}_3=
\Nc C_F \frac{C_A^2}{4}\:,
\vspaceinarray \\ &&
\widetilde{T}_3^\dag \widetilde{T}_4 = \Nc C_F \frac{C_A^2}{2}\:.
\eeeq
Using these identities the square of the matrix element can be written
in the form:
\beeq
&&
\left|{\cal M}_5(1_q, 2_\qb, 3_\gt, 4_\gt, 5_g)\right|^2 =
\\ \nn && \qquad
\Nc C_F^3\left\{ x\widetilde{M}_x + x^2 \widetilde{M}_{xx}\right\}\:,
\eeeq
where
\beeq
&&
\widetilde{M}_x = \left|\widetilde{M}_1 + \widetilde{M}_2\right|^2\:,
\vspaceinarray \\ &&
\widetilde{M}_{xx} = \left|\widetilde{M}_3 + \widetilde{M}_4\right|^2 
\\ \nn && \qquad\:
+ \frac12 {\rm Re}
\big((\widetilde{M}_1 + \widetilde{M}_2)
     (\widetilde{M}_3 + \widetilde{M}_4)^*\big)                    
\\ \nn && \qquad\:
-{\rm Re}
\big((\widetilde{M}_1 + \widetilde{M}_4)
     (\widetilde{M}_2 + \widetilde{M}_3)^*\big)\:.
\eeeq

\def\ibid#1#2#3  {{\it ibid} {\bf #1}, #2 (19#3)}
\def\np#1#2#3  {Nucl.\ Phys.\ {\bf #1}, #2 (19#3)}
\def\npproc#1#2#3  {Nucl.\ Phys.\ B (Proc.\ Suppl.) {\bf #1}, #2 (19#3)}
\def\plb#1#2#3  {Phys.\ Lett.\ B {\bf #1}, #2 (19#3)}
\def\pl#1#2#3  {Phys.\ Lett.\ {\bf #1}, #2 (19#3)}
\def\prep#1#2#3  {Phys.\ Rep.\ {\bf #1}, #2 (19#3)}
\def\prd#1#2#3 {Phys.\ Rev.\ D {\bf #1}, #2 (19#3)}
\def\prl#1#2#3 {Phys.\ Rev.\ Lett.\ {\bf #1}, #2 (19#3)}
\def\zpc#1#2#3  {Zeit.\ Phys.\ C {\bf #1}, #2 (19#3)}
\def\cpc#1#2#3  {Comp.\ Phys.\ Comm.\ {\bf #1}, #2 (19#3)}
\def\ncim#1#2#3 {Nuovo Cim.\ {\bf #1}, #2 (19#3)}
\def\nciml#1#2#3 {Nuovo Cim.\ Lett.\ {\bf #1}, #2 (19#3)}
\def\jhep#1#2#3 {J. of High Energy Phys.\ {\bf #1}, #2 (19#3)}
\def\etal{{\em et al.}}


\begin{thebibliography}{99}
\bibitem{QCD}
M. Gell-Mann, Acta Phys.\ Aus., Suppl.\ {\bf IX}, 733 (1972);
H. Fritzsch and M. Gell-Mann, XVIth International Conference on High Energy
Physics, Vol. II p.135 (1972);
H. Fritzsch, M. Gell-Mann and H. Leutwyler, \pl{47B}{365}{73} ;
T. Muta, Foundation of Quantum Chromodynamics, World Scientific, (1987).
%\bibitem{QCDtests}
\bibitem{KEK}
VENUS collaboration, K. Abe \etal, \plb{240}{232}{90} ;
AMY collaboration, K.B. Lee \etal, \ibid{313}{469}{93} ;
TOPAZ collaboration, Y. Ohnishi \etal, \ibid{313}{475}{93} ;
\bibitem{SLAC}
Mark II collaboration, S. Komamiya \etal, \prl{64}{987}{90} ;
SLD collaboration, K. Abe \etal, \prd{51}{962}{95} .
\bibitem{ALEPH}
ALEPH collaboration, D. Decamp \etal, \plb{257}{479}{91} ;
\ibid{284}{163}{92} .
\bibitem{DELPHI}
DELPHI collaboration, P. Abreu \etal, \zpc{54}{55}{92} ;
\ibid{59}{21}{93} .
\bibitem{OPAL}
OPAL collaboration, P.D. Acton \etal, \zpc{55}{1}{92} ; \ibid{59}{93}{1} ;
R. Akers \etal, \ibid{68}{519}{95} .
\bibitem{L3}
L3 collaboration, O. Adriani \etal, \plb{284}{471}{92} ; \prep{236}{1}{93} .
\bibitem{lg}
G.R. Farrar, \plb{265}{395}{91} ; \prd{51}{3904}{95} ; hep-ph/9704309;
J. Ellis, D. Nanopoulos and D. Ross, \plb{305}{375}{93} ;
R. Mu$\tilde{\rm n}$oz-Tapia and W.J. Stirling, \prd{49}{3763}{94} .
F.E. Close, G.R. Farrar and Z.P. Li, \ibid{55}{5749}{97} ;
D.J. Chung, G.R. Farrar and E.W. Kolb, astro-ph/9707036;
\bibitem{Cmeasure}
%\bibitem{L3}
L3 Collaboration, B. Adeva \etal, \plb{248}{227}{90} ;
%\bibitem{a34}
DELPHI Collaboration, P. Abreu \etal, \ibid{255}{466}{91} ;
%\bibitem{DELPHI}
DELPHI Collaboration, P. Abreu \etal, \zpc{59}{357}{93} ;
%\bibitem{OPAL}
OPAL Collaboration, R. Akers \etal, \ibid{65}{367}{95} ;
%\bibitem{ALEPH}
ALEPH Collaboration, D. Decamp \etal, \plb{284}{151}{92} ;
ALEPH Collaboration, R. Barate \etal, \zpc{76}{1}{97} .
\bibitem{SDjets}
A. Signer and L. Dixon, \prl{78}{811}{97} ;
\bibitem{DSjets}
L. Dixon and A. Signer, \prd{56}{4031}{97} .
\bibitem{Signer} A. Signer, hep-ph/9705218.
\bibitem{NTDpar}
Z. Nagy and Z. Tr\'ocs\'anyi, \prl{79}{3604}{97} .
\bibitem{NTFox}
Z. Nagy and Z. Tr\'ocs\'anyi, \npproc{64}{63}{98} .
\bibitem{NTangulars}
Z. Nagy and Z. Tr\'ocs\'anyi, \prd{57}{5793}{98} .%hep-ph/9712385.
\bibitem{Glover} E.W.N. Glover, hep-ph/9805481.
\bibitem{GM4q} E.W.N. Glover and D.J. Miller, \plb{396}{257}{97} .
\bibitem{CGM2q2g} J.M. Campbell, E.W.N. Glover and D.J. Miller, 
\plb{409}{503}{97} .
\bibitem{BDKW4q} Z. Bern, L. Dixon, D. A. Kosower and S. Wienzierl,
\np{B489}{3}{97} ;
\bibitem{BDK2q2g} Z. Bern, L. Dixon and D. A. Kosower,
\np{B513}{3}{98} .
\bibitem{a5parton}
K. Hagiwara and D. Zeppenfeld, \np{B313}{560}{89} ;
F.A. Berends, W.T. Giele and H. Kuijf, \ibid{B321}{39}{89} ;
N.K. Falk, D. Graudenz and G. Kramer, \ibid{B328}{317}{89} .
\bibitem{durham} S. Catani Yu.L. Dokshitzer, M. Olsson, G. Turnock and
B.R. Webber, \plb{269}{432}{91} .
\bibitem{cambridge}
Yu.L. Dokshitser, G.D. Leder, S. Moretti, B.R. Webber, \jhep{8}{1}{97} .
\bibitem{dimreg}
G. 't Hooft and M. Veltman, \np{B44}{189}{72} ;
G. Bollini and J.J. Giambiagi, \ncim{12B}{20}{72} ;
J.F. Ashmore, \nciml{4}{289}{72} ;
G.M. Cicuta and E. Montaldi, \ibid{4}{329}{72} ;
R. Gastmans and R. Meuldermans, \np{B63}{277}{73} .
\bibitem{CSdipole}
S. Catani and M.H. Seymour, \plb{378}{287}{96} ; \np{B485}{291}{97} .
\bibitem{NTloopamps} Z. Nagy and Z. Tr\'ocs\'anyi, \plb{414}{187}{97} .
\bibitem{DSpriv}
L. Dixon and A. Signer, private communication.
%\bibitem{JADE} JADE Collaboration, W. Bartel \etal, \zpc{33}{23}{86} .
\bibitem{cambridgeB} S. Bentvelsen and I. Meyer, hep-ph/9803322 .
\bibitem{CDOTW} S. Catani, Yu.L. Dokshitzer, M. Olsson, G. Turnock and
B.R. Webber, \plb{269}{432}{91} .
\bibitem{DSansatz}
G. Dissertori and M. Schmelling, \plb{361}{167}{95} .
\bibitem{CMW} S. Catani, G. Marchesini and B.R. Webber, \np{B349}{635}{91} .
\bibitem{Kcoeff}
J. Kodaira and L. Trentadue, \pl{112B}{66}{82} ;
C.T.H. Davies, W.J. Stirling and B.R. Webber, \np{B256}{413}{85} ;
S. Catani, E. d'Emilio and L. Trentadue, \pl{211B}{335}{88} .
\bibitem{Catani} S. Catani,
in QCD at 200\,TeV, Proc.\ 17th INFN Eloisatron Project Workshop,
ed.: L. Cifarelli and Yu.L. Dokshitzer, Plenum Press, New York (1992).
\bibitem{ALEPHR4} ALEPH collaboration, R. Barate \etal, \prep{294}{1}{98} .
\bibitem{PYTHIA} T. Sj\"ostrand, \cpc{82}{74}{94} .
\bibitem{HERWIG} 
G. Marchesini and B.R. Webber, \np{B310}{461}{88} ;
G. Marchesini, B.R. Webber, G. Abbiendi, I.G. Knowles, M.H. Seymour and
L. Stanco, \cpc{67}{465}{92} .
\bibitem{LEPshapes} 
%\bibitem{OPALshapes}
OPAL Collaboration, M.Z. Akrawy \etal, \zpc{47}{505}{90} ;
G. Alexander \etal, \ibid{72}{191}{96} ;
%\bibitem{L3shapes}
L3 Collaboration, B. Adeva \etal, \ibid{55}{39}{92} ;
%\bibitem{DELPHIshapes}
DELPHI Collaboration, P. Abreu \etal, \ibid{73}{11}{96} .
\bibitem{Tmin} MARKJ Collaboration, D.P. Barber \etal, \prl{43}{830}{79} .
\bibitem{Dpar}                                                 
G. Parisi, \pl{74B}{65}{78} ;                
J.F. Donoghue, F.E. Low and S.Y. Pi, \prd{20}{2759}{79} .
\bibitem{KNMW}                                                 
%\bibitem{ERT}
R.K. Ellis, D.A. Ross and A.E. Terrano, \np{B178}{421}{81} ;
Z. Kunszt, P. Nason, G. Marchesini and B.R. Webber, in `Z Physics at LEP
1', CERN 89-08, vol.\ 1, p.\ 373.
\bibitem{debrecen} See the URL http://dtp.atomki.hu/HEP/pQCD.
\bibitem{BGcurrents}
F.A. Berends and W.T. Giele, \np{B306}{759}{88} .
\bibitem{helicity}
P. de Causmaecker, R. Gastmans, W. Troost and T.T. Wu, \pl{105B}{215}{81} ;
R. Kleiss, \np{B241}{61}{84} ;
F.A. Berends, P.H. Daverveldt and R. Kleiss, \ibid{253}{441}{85} ;
J.F. Gunion and Z. Kunszt, \pl{161B}{333}{85} ;
Z. Xu, D.H. Zhang and L. Chang, \np{B291}{392}{87} .
\bibitem{MP}
M.L. Mangano and S.J. Parke, \prep{200}{301}{91} , and references therein.
\end{thebibliography}
\end{document}